\documentclass[iop]{emulateapj}
\usepackage{color}

\def\gsim{\mathrel{\raise0.35ex\hbox{$\scriptstyle >$}\kern-0.6em 
\lower0.40ex\hbox{{$\scriptstyle \sim$}}}}
\def\lsim{\mathrel{\raise0.35ex\hbox{$\scriptstyle <$}\kern-0.6em 
\lower0.40ex\hbox{{$\scriptstyle \sim$}}}}

\def\nii{{\rm [N{\sc ii}]}}

\slugcomment{ Accepted for publication in ApJ}

\shorttitle{Star formation main sequence around $z\sim 1.5$ cluster}
\shortauthors{Y. Koyama et al.}

\begin{document}


\title{The environmental impacts on the star formation main sequence:\\ 
an H$\alpha$ study of the newly discovered rich cluster at z=1.52}


\author{Yusei Koyama$^{1,2}$, 
Tadayuki Kodama$^{1,3}$, 
Ken-ichi Tadaki$^{1}$, 
Masao Hayashi$^{1,4}$, \\
Ichi Tanaka$^{5}$, 
and Rhythm Shimakawa$^{3}$}

\email{koyama.yusei@nao.ac.jp}


\altaffiltext{1}{National Astronomical Observatory of Japan, Mitaka, Tokyo 181-8588, Japan}
\altaffiltext{2}{Institute of Space Astronomical Science, Japan Aerospace Exploration Agency, Sagamihara, Kanagawa 252-5210, Japan}
\altaffiltext{3}{Department of Astronomical Science, The Graduate University for Advanced Studies, Mitaka, Tokyo 181-8588, Japan}
\altaffiltext{4}{Institute for Cosmic Ray Research, The University of Tokyo, Kashiwa, Chiba 277-8582, Japan}
\altaffiltext{5}{Subaru Telescope, National Astronomical Observatory of Japan, 650  North A'ohoku Place, Hilo, HI 96720, USA}


\begin{abstract}
We report the discovery of a strong over-density of galaxies in the field of a radio galaxy at $z=1.52$ (4C65.22) based on our broad-band and narrow-band (H$\alpha$) photometry with Subaru Telescope. We find that H$\alpha$ emitters are located in the outskirts of the density peak (cluster core) dominated by passive red-sequence galaxies. This resembles the situation in lower-redshift clusters, suggesting that the newly discovered structure is a well-evolved rich galaxy cluster at $z=1.5$. Our data suggest that the color--density and stellar mass--density relations are already in place at $z\sim 1.5$, mostly driven by the passive red massive galaxies residing within $r_c$$\lsim$200~kpc from the cluster core. These environmental trends almost disappear when we consider only star-forming (SF) galaxies. We do not find SFR--density or SSFR--density relations amongst SF galaxies, and the location of the SF main sequence does not significantly change with environment. Nevertheless, we find a tentative hint that star-bursting galaxies (up-scattered objects from the main sequence) are preferentially located in a small group at $\sim$1-Mpc away from the main body of the cluster. We also argue that the scatter of the SF main sequence could be dependent on the distance to the nearest neighboring galaxy.

\end{abstract}





\section{Introduction}

Recent observations have established that star-forming (SF) galaxies show a tight correlation between stellar mass ($M_{\star}$) and star-formation rate (SFR). This SFR--$M_{\star}$ correlation (so-called SF main sequence) is investigated in the local universe (e.g.\ \citealt{bri04}; \citealt{pen10}), as well as in the distant universe out to $z\gsim 2$ (e.g.\ \citealt{noe07}; \citealt{dad07}; \citealt{elb07}; \citealt{san09}; \citealt{kaj10}; \citealt{rod10}; \citealt{bau11b}; \citealt{whi12}). The SFR at a fixed mass evolves with time, reflecting the cosmic star formation history. The presence of this tight SFR--$M_{\star}$ correlation suggests that stellar mass has always been an important parameter that regulates SF activity in galaxies across cosmic time.

The scatter around the SF main sequence is reported to be very small ($\lsim$0.3~dex level at any redshifts), but the deviation around the SF main sequence may be an important parameter as it is likely to reflect the variation of gas accretion/exaustion history of galaxies (e.g.\ \citealt{elb11}; \citealt{sai12}; \citealt{tac13}). Some recent studies attempt to identify the origin of the scatter and the key parameter that makes the strongest impact on galaxy evolution. For example, \cite{wuy11} studied how the galaxy structure and the mode of SF activity depend on the position on the SFR--$M_{\star}$ diagram. They find that the upper envelope of the SF main sequence tends to be dominated by dusty galaxies (characterized by high SFR$_{\rm IR}$/SFR$_{\rm UV}$ ratio) with high sersic index ({\it n}), suggesting a rapid build-up of mass in the nuclear regions of these systems, due to e.g. galaxy-galaxy interactions/mergers.

\begin{table*}
\begin{center}
\caption{Summary of our Subaru data of the 4C65.22 field. \label{tbl-1}}
\begin{tabular}{ccccccc}
\hline 
\hline
Filter & Instrument & Obs. date  & Exp time   & PSF size & Limit mag.   & $A_{\rm band}$     \\
       &            &   [UT]     & [min]      & [arcsec] & (1.5$''$, 5$\sigma$, AB) & [mag]       \\
\hline 
$B$    & Suprime-Cam  & 30 Aug 2005    &  108  & 0.95  & 27.4  &  0.19 \\
$r'$   & Suprime-Cam  & 7,8 Oct 2010  &   94  & 0.92  & 26.9  &  0.12  \\
$z'$   & Suprime-Cam  & 6 Oct 2010    &   42  & 0.62  & 25.3  &  0.06  \\
$J$    & MOIRCS       & 6 Sep 2011     &   87  & 0.60  & 23.9  &  0.04  \\
$H$    & MOIRCS       & 6,7 Sep 2011   &   66  & 0.49  & 23.3  &  0.03 \\
$K_s$  & MOIRCS       & 8 Sep 2011     &   48  & 0.60  & 23.2  &  0.02 \\
NB1657 & MOIRCS       & 7,8 Sep 2011   &  183  & 0.56  & 22.6  &  0.03 \\
\hline
\end{tabular}
\vspace{3mm}
\tablecomments{The $B$-band data are retrieved from SMOKA (Subaru Science Archive: http://smoka.nao.ac.jp/). }
\end{center}
\end{table*}

Another important parameter that could bring strong impacts on SF activity of galaxies is ``environment''. The morphology--density or color--density relation is widely recognized in the local universe (e.g.\ \citealt{dre80}; \citealt{lew02}; \citealt{gom03}; \citealt{got03}; \citealt{tan04}). However, if we focus on the star-forming population, their properties are not necessarily strong functions of environment. \cite{bal04} used SDSS and 2dFGRS dataset to show the H$\alpha$ equivalent width (EW) amongst SF galaxies is independent of environment (see also \citealt{wij12}). \cite{pen10} used local SF galaxies drawn from SDSS to show that the SF main sequence is indistinguishable between high- and low-density environment. They argue that the environment does change the ``fraction'' of SF galaxies, while it has very little impact on the SF main sequence.

An observational challenge here is to test if the ``universality'' of SF main sequence holds in the distant universe, where the average star-formation activity is about an order of magnitude higher (e.g.\ \citealt{mad96}). Some recent studies have attempted to identify the environmental dependence of the SF main sequence from intermediate- to high-redshift universe out to $z\sim 2$ (\citealt{vul10}; \citealt{li11}; \citealt{mcg11}; \citealt{muz12}; \citealt{gre12}; \citealt{tan11}; \citealt{gru11}; \citealt{koy13a}; \citealt{zei13}), but a full consensus has not yet been obtained, because of the different sample selection and/or different environment definitions. The most recent study by \cite{lin14} used a large sample of galaxies drawn from Pan-STARRS, and demonstrated that the (S)SFR--$M_{\star}$ relation for SF galaxies is comparable between field and group environment out to $z\sim 0.8$. They also reported that there is a moderate ($\sim$17\%) SFR decrease for SF galaxies (at a given mass) in cluster environment. Pushing this kind of study toward the critical epoch of galaxy formation (i.e.\ $z$$\sim$1--3) is clearly an important step, but constructing a large uniformly selected SF galaxy sample across environment at such high-redshift universe has been very challenging.

Recently, \cite{koy13b} made an important step on this issue. They compiled a large, H$\alpha$-selected SF galaxy sample in distant clusters (from MAHALO-Subaru project; \citealt{kod13}) and in general field environments at $z=0.4,0.8,2.2$ (from HiZELS; \citealt{sob13}), and demonstrated that the environmental impacts on the SF main sequence are likely to be always small since $z\sim 2$ ($<$0.2~dex level at maximum), as far as we rely on the simple H$\alpha$-based SFRs. However, in contrast to this apparently simple picture, we also find a tentative hint that SF galaxies in distant cluster environments tend to be more massive (see also \citealt{lin14}), and perhaps more highly obscured by dust. In this paper, we provide a more detailed look on this issue, and attempt to identify the environmental impacts on the SF main sequence, using our newly discovered rich cluster field at $z\sim 1.5$. 

The paper is organized as follows. In \S~2, we present our Subaru data of the 4C\,65.22 field. After selecting $z\approx 1.5$ galaxies based on our photometric data in \S~3, we report the discovery of a strong over-density of $z\sim 1.5$ galaxies near the radio galaxy in \S~4. In \S~5, we present environmental dependence of galaxy properties at $z\sim 1.5$ across the newly discovered structure. In \S~6, we discuss our results in line with recent studies. Finally, our conclusions are given in \S~7. Throughout the paper, we adopt $\Omega_{\rm{M}} =0.3$, $\Omega_{\Lambda} =0.7$, and $H_0 =70$ km s$^{-1}$Mpc$^{-1}$, which gives a 1$''$ scale of 8.46 kpc and the cosmic age of 4.2~Gyr at the redshift of our target cluster associated to the radio galaxy, 4C65.22 ($z=1.520$). Magnitudes are all given in the AB system. \\

\section{Observation and Data Reduction}

\subsection{The field of a radio galaxy 4C\,65.22 (z=1.52)}

In this paper, we aim to present our newly discovered rich galaxy cluster at $z\sim 1.5$. An increasing number of high-$z$ galaxy clusters are now being discovered with various techniques (e.g.\ \citealt{tan10}; \citealt{pap10}; \citealt{fas11}; \citealt{gob11}; \citealt{new13}), but it is still very important to increase the sample of clusters/groups in the distant universe. For this purpose, high-$z$ radio galaxies (HzRGs) are often used as "landmarks" for high-$z$ (proto-)clusters since HzRGs are expected to be the progenitors of present-day massive cluster ellipticals (e.g.\ \citealt{kur00}; 2004; \citealt{ven02}; \citealt{bes03}; \citealt{kaj06}; \citealt{ven07}; \citealt{kod07}; \citealt{gal10}; 2013).  These studies have been successful in identifying prominent high-$z$ structures. 

We have been undertaking an intensive observational program of high-$z$ star-forming galaxies using narrow-band filters on the Subaru Telescope, {\it MApping HAlpha and Lines of Oxygen with Subaru} (MAHALO-Subaru; \citealt{kod13}, \citealt{tad11}; \citealt{hay12}; \citealt{koy13a}; \citealt{hay14}). As a part of this observational campaign, we target a radio galaxy field, 4C\,65.22, at $z=1.520$ ($\alpha =17^{h}47^{m}13^{s}.9$, $\delta = +65^{\circ}32'36''$ in J2000)\footnote{We exploit $z=1.520$ as the redshift of 4C65.22 following the [O{\sc ii}]-line-based measurement presented by \cite{kol95}. These authors reported slightly different redshifts for other line measurements (such as C$_{\rm IV}$, C$_{\rm III]}$, or Mg$_{\rm II}$), although the difference is negligibly small for our study ($\Delta z \lsim$0.005 level).}, which is located near the North Ecliptic Pole (NEP). The 4C65.22 is identical to 8C\,1747+655, and a strong X-ray detection is reported from this radio galaxy (see \citealt{bri99}; \citealt{kol94}; 1995; \citealt{lay93}). Below we describe our new Subaru observations of this radio galaxy field.

\subsection{Near-infrared data with MOIRCS}

\begin{figure}
 \begin{center}
 \vspace{0.2cm}
 \rotatebox{270}{\includegraphics[height=7.7cm]{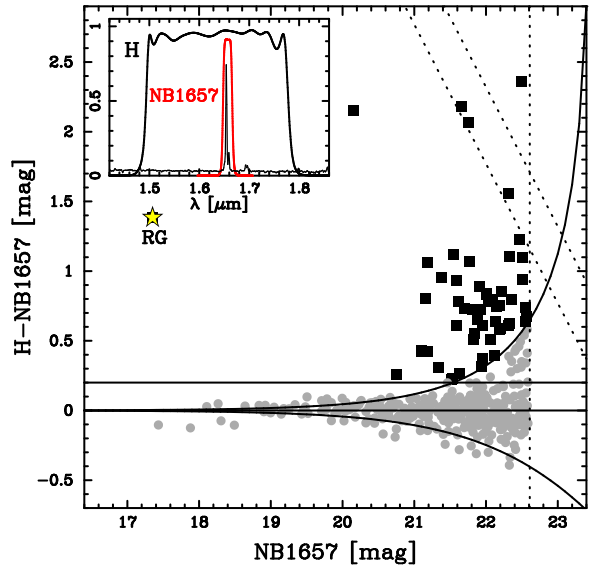}}
 \end{center}
 \vspace{-0.3cm}
\caption{ Selection of the NB1657 emitters. We plot $H-$NB1657 colors of all NB-selected galaxies against their NB magnitudes. The vertical dotted line shows the 5$\sigma$ limiting magnitude in NB1657, and the slanted dotted-lines show 2- and 3-$\sigma$ limits in $H$-band. The solid-line curves show $\pm$2$\Sigma$ excesses in the $H-$NB1657 color. We select galaxies with $H-$NB1657$>$0.2 and with $H-$NB1657$>$2$\Sigma$ as NB1657 emitters (black squares). The brightest galaxy (marked with yellow star) is the central radio galaxy (4C\,65.22), showing a strong NB excess as expected. In the inset, we show the transmission curves of the broad-band ($H$) and narrow-band (NB1657) filters. Also shown is the Sc-type galaxy spectrum (redshifted to $z=1.52$) from Kinney et al. (1996).\\ 
\label{fig:emitter_select}}
\end{figure}

We observed the 4C\,65.22 field with broad-band ($JHKs$) and narrow-band (NB1657) filters using Multi-Object InfraRed Camera and Spectrograph (MOIRCS; \citealt{ich06}, \citealt{suz08}) on the Subaru Telescope (\citealt{iye04}). The MOIRCS covers a 4$'\times$7$'$ field of view (FoV), which corresponds to the physical scale of 2.0$\times$3.6 Mpc$^2$ at $z\sim 1.5$. We note that the NB1657 filter ($\lambda_c=$1.657$\mu$m, $\Delta \lambda =$0.019$\mu$m; see Fig.~\ref{fig:emitter_select}) captures the H$\alpha$ lines at $z=1.52^{+0.02}_{-0.01}$ (or equivalently $-1200\lsim \Delta v \lsim 2300$ km\,s$^{-1}$ from the radio galaxy), and this filter plays a critical role in this study (see \S~3.1 for the H$\alpha$ emitter selection). The observations were carried out in September 2011 under excellent conditions. The data were reduced using the MCSRED software (\citealt{tan11}) in a standard manner. For the NB data reduction, we use the "{\sc nbmcsall}" task provided in the MCSRED package, which is dedicated to the MOIRCS NB data reduction. This task takes a careful treatment for the variation of sky patterns over the MOIRCS FoV (see \citealt{tan11} for details). The photometric zero points are derived with the standard stars observed during the same observing nights. The exposure times, PSF sizes, and limiting magnitudes of our MOIRCS data are summarized in Table~1.

\subsection{Optical data with Suprime-Cam}

We also use {\it B}, $r'$, and $z'$-band data of the 4C\,65.22 field taken with the Prime Focus Camera on Subaru (Suprime-Cam; \citealt{miy02}). The $r'$ and $z'$-band data are newly obtained by our team in October 2010, while the {\it B}-band data are retrieved from Subaru Science Archive (SMOKA). All the data are reduced with SDFRED pipeline software developed for the Suprime-Cam data reduction (\citealt{yag02}; \citealt{ouc04}). We use SDFRED1 for the {\it B}-band data reduction, while we use SDFRED2 (minor updated version) for the reduction of $r'$/$z'$ data according to the Suprime-Cam CCD upgrade in 2008. The magnitude zero points are derived from the data of standard stars taken during the same observing nights. We note that the original field of view (FoV) of Suprime-Cam is 34$'\times$27$'$, but we only use central 4$'\times$7$'$ region in this study due to the limited FoV of our near-infrared (NIR) data. The final optical images are aligned to our MOIRCS image and re-gridded to the pixel scale same as the MOIRCS data (0.117$''$/pix). 

\subsection{Photometric catalog}

The photometric catalog is created with SExtractor ver.2.5.0 (\citealt{ber96}). Before running SExtractor, we apply gaussian smoothing for the NIR images, and match their PSF sizes to 0.62$''$ (i.e.\ PSF of $z'$-band image). We note that the $B$- and $r'$-band data have broader PSF size (0.9$''$--1.0$''$; see Table~1). We apply aperture correction of 0.165/0.145 mag for $B$/$r'$-band data when we perform aperture photometry, rather than degrading our good-quality NIR data by $\sim$50--100\%. These correction values are determined by a simple simulation using our $z'$-band image: we smoothed the $z'$-band image to 0.95$''$/0.92$''$ (i.e.\ the PSF size of $B$/$r'$-band image), and then performed 1.5$''$ aperture photometry of the same sources on the smoothed/unsmoothed images to determine the typical fluxes escaped from the aperture.

The limiting magnitudes are determined by measuring the deviation of 1.5$''$ aperture photometry (2.5$\times$PSF size) at random positions over the FoV, and the measured limiting magnitudes are reported in Table~1. The measured 1-$\sigma$ random sky noise on each image is also used as the photometric uncertainties of individual sources. Following our previous high-$z$ emission-line galaxy surveys carried out with the MAHALO project, we construct a NB-selected source catalog using the double image mode of SExtractor. In this study, we use 1.5$''$ aperture magnitudes ({\sc mag\_aper}) for source detection and measuring colors, while we use total magnitude ({\sc mag\_auto}) when we derive physical quantities such as $M_{\star}$ or SFRs of galaxies. 

\begin{figure*}
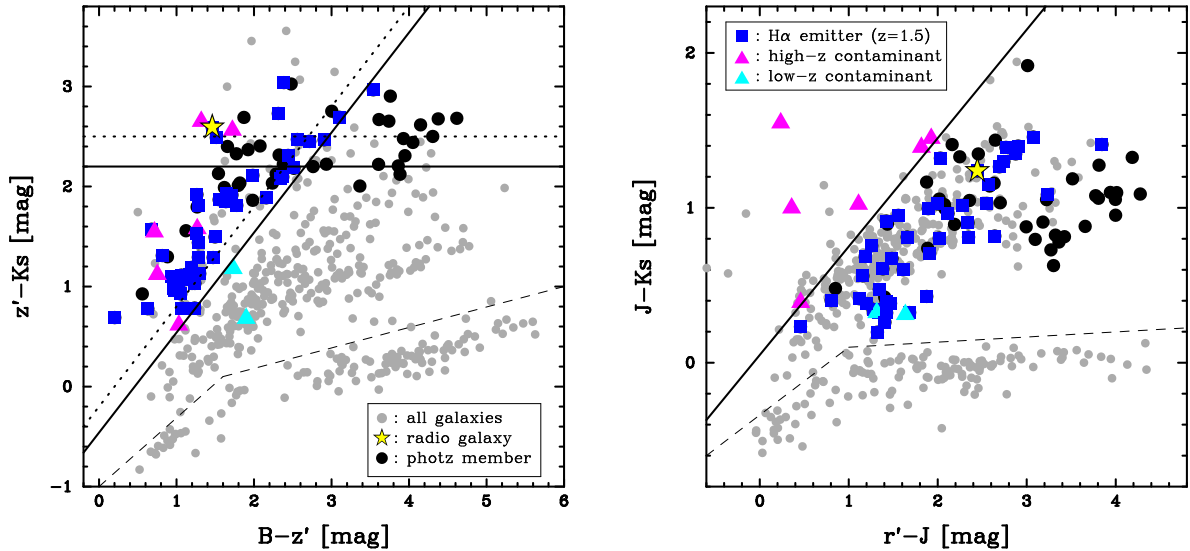

 \begin{center}
 \vspace{0.2cm}
 \includegraphics[angle=270,scale=0.43]{fig2a.ps}
 \hspace{7mm}\includegraphics[angle=270,scale=0.43]{fig2b.ps}
 \end{center}
 \vspace{-0.2cm}
\caption{ The color selections of H$\alpha$ emitters at $z=1.52$ based on the {\it BzK} (left) and {\it rJK} (right) color--color diagrams. We here plot all NB-selected sources with $>$2$\sigma$ detection at all broad-band filters including galactic stars (grey dots). The NB1657 emitters selected in Fig.~1 are shown with colored symbols. The {\it BzK} diagram isolates low-$z$ contamination (cyan triangles), and the {\it rJK} diagram isolates high-$z$ contamination such as [O{\sc iii}] emitters at $z=2.31$ (magenta triangles). The H$\alpha$ emitters are defined as NB1657 emitters satisfying both {\it BzK} and {\it rJK} criteria, and they are shown with blue squares. The solid lines show the criteria applied in this study. The dotted lines in the left panel show the original {\it BzK} criteria proposed by Daddi et al. (2004), while we apply the modified-{\it BzK} criteria to improve completeness for $z\sim 1.5$ galaxies (Hayashi et al. 2011). Sources below the thin-dashed lines are galactic stars and discarded from our catalog. The radio galaxy (shown with yellow star) satisfies both {\it BzK} and {\it rJK} criteria. We also plot photo-$z$ selected non-emitters with black circles ($1.3<z_{\rm phot}<1.7$; see \S~3.2). As expected, most of the photo-$z$ members satisfy both color criteria. \\
\label{fig:color_select}}
\end{figure*}

Using the stellar spectral templates of \cite{gun83}, we checked colors of stellar objects within the FoV. We find that the derived zero points show good agreement with the template, but we apply a small zero point correction to $H$-band magnitudes by $+$0.1~mag to be consistent with the stellar template. After removing saturated objects and galactic stars (based on the $BzK$ and $rJK$ colors; see Fig.~\ref{fig:color_select}), our final catalog includes 393 objects down to $m_{\rm NB}=$22.6~mag (corresponding to 5$\sigma$ limit in NB1657). At the position of 4C\,65.22, we estimate the dust reddening to be $E$$(B-V)$$=$0.042~mag based on the dust map of \cite{sch98}. We derive the correction value at each band using the extinction law of \cite{car89}, and the derived correction values ($A_{\rm band}$) are listed in Table~1. \\

\section{Sample selection}

In this section, we select galaxies at $z\approx 1.5$ using our photometric catalog. The primary goal of this project is to map the H$\alpha$ emitting galaxies around 4C65.22 with the narrow-band technique (\S~3.1). We also apply photometric redshifts (photo-$z$) to select more quiescent galaxy population (\S~3.2). By combining these star-forming and quiescent galaxies, we construct a catalog of "member" galaxies associated to the central radio galaxy.

\subsection{H$\alpha$ emitter selection}

We first select H$\alpha$ emitters at $z$$=$1.52 using the narrow-band excess technique. Our selection includes the following two steps: (1) identification of NB excess galaxies using $H-$NB1657 color, and (2) removal of contaminant emitters (such as [O{\sc ii}]/H$\beta$/[O{\sc iii}] emitters at higher redshifts, or various line-emitting galaxies at low redshifts) based on their broad-band colors. 

In Fig.\,\ref{fig:emitter_select}, we plot $H-$NB1657 colors of all NB-selected galaxies against their NB magnitudes. To select NB excess sources, we apply $H-$NB1657$>$0.2 and $H-$NB1657$>$2$\Sigma$, where $\Sigma$ denotes the significance of NB excess (\citealt{bun95}). This yields 54 galaxies with NB excess, and they are shown with the black squares in Fig.\,\ref{fig:emitter_select}. We note that the former condition corresponds to EW$_{\rm rest}$$\gsim$20\AA\, and the latter corresponds to dust-free SFR$_{\rm H\alpha}$ of $\gsim$3$M_{\odot}$/yr (see more details in \S~5.2). 

Next, we remove the expected contaminant galaxies, such as low-redshift Pa$\beta$ emitters at $z\approx 0.3$, H$\beta$/[O{\sc iii}] emitters at $z\approx 2.3$, or [O{\sc ii}] emitters at $z\approx 3.3$. The presence of a strong emission line located at $\lambda \simeq 1.657\mu$m allows us to determine their redshifts using broad-band color information (see e.g. \citealt{kod04}; \citealt{koy10}). We find that most of the NB1657 emitters are detected with our deep broad-band images, while we find two of the NB1657 emitters are fainter than the 2$\sigma$ limit in $K_s$-band. The $K_s$-band photometry is very important in the following member selection, and it is also essential for deriving stellar mass of galaxies (see \S~5.3). We therefore exclude these two $K_s$-undetected objects from the following analysis. 

In Fig.~\ref{fig:color_select} (left), we show the {\it BzK} diagram to remove low-$z$ contamination. The $BzK$ method is originally proposed by Daddi et al. (2004), and is one of the most commonly used color selections for high-$z$ galaxies. The {\it BzK} method is designed to select galaxies at $1.4\lsim z\lsim 2.5$ and optimized for $z\sim 2$ galaxies. Therefore galaxies at $z\sim 1.5$ are usually located near the boarder lines of the {\it BzK} criteria. \cite{hay10} proposed a "modified {\it BzK} selection" for selecting $z\sim 1.5$ galaxies, which slightly softens the criteria toward the bluer side by $\sim$0.2 magnitude (see solid lines in Fig.~\ref{fig:color_select} left). The robustness of this method is demonstrated by our spectroscopic follow-up study (\citealt{hay11}). The colored symbols in Fig.~\ref{fig:color_select} show the NB1657 emitters selected above. We find that most of the emitters satisfy the {\it sBzK} criteria as expected, while two of the NB emitters are likely low-$z$ contamination (shown as cyan triangles in Fig.~\ref{fig:color_select}). 

We then apply {\it rJK} selection to remove high-$z$ contamination at $z>2$ (the right-hand panel of Fig.~\ref{fig:color_select}). This method is proposed by \cite{hay12} to distinguish $z\gsim 2.5$ galaxies from $z<2$ galaxies, and the robustness of this method is shown by our recent spectroscopic observation (\citealt{shi14}). With this technique, we find that six of the NB1657 emitters are likely high-$z$ emitters at $z>2$ (see magenta triangles in Fig.~\ref{fig:color_select}), and they are now excluded from our catalog. In summary, among the 52 NB emitter candidates (with $K_s$-band detection), we identify total eight contaminant emitters (two low-$z$ and six high-$z$ objects). In the following analysis, we consider the remaining 44 emitters as H$\alpha$ emitters at $z=1.52$. The list of the selected H$\alpha$ emitters is provided in Table~2.

\begin{table*}
\begin{center}
\caption{List of the 44 H$\alpha$ emitters in the field of 4C\,65.22 selected in \S~3.1.} \label{tbl-2}
\begin{tabular}{lllccccc}
\hline 
\hline
ID & R.A. & Dec. & $H$~mag. & NB1657~mag. & EW$_{\rm r}$(H$\alpha$+[N{\sc ii}])  & $F$(H$\alpha$+[N{\sc ii}]) & $\log(M_{\star}/M_{\odot})$\tablenotemark{\ddag}  \\
   & [deg] & [deg] & [AB, 1.5$''$] & [AB, 1.5$''$] & [\AA] & [10$^{-16}$ erg/s/cm$^2$] &    \\
\hline 
12 & 266.76831 & 65.57383 & $22.57\pm0.10$ & $21.96\pm0.11$ & $83.3\pm6.3$ & $1.53\pm0.35$ & $10.43\pm0.06$ \\
13 & 266.76767 & 65.57382 & $22.96\pm0.14$ & $22.33\pm0.16$ & $11.4\pm6.2$ & $0.13\pm0.25$ & $10.41\pm0.05$ \\
14 & 266.67044 & 65.57362 & $23.82\pm0.30$ & $21.75\pm0.09$ & $1069.6\pm274.2$ & $1.73\pm0.24$ & $9.82\pm0.11$ \\
293 & 266.82201 & 65.51409 & $22.65\pm0.11$ & $21.92\pm0.11$ & $163.1\pm12.8$ & $1.42\pm0.27$ & $10.14\pm0.08$ \\
357 & 266.76742 & 65.51608 & $22.53\pm0.10$ & $21.88\pm0.10$ & $94.1\pm8.4$ & $0.93\pm0.25$ & $10.75\pm0.04$ \\
409 & 266.78674 & 65.51795 & $22.33\pm0.08$ & $21.95\pm0.11$ & $50.0\pm5.9$ & $0.64\pm0.25$ & $10.70\pm0.03$ \\
576 & 266.70583 & 65.52318 & $21.89\pm0.06$ & $21.62\pm0.08$ & $21.6\pm3.5$ & $0.45\pm0.24$ & $11.06\pm0.02$ \\
761 & 266.89357 & 65.53110 & $22.89\pm0.14$ & $22.14\pm0.13$ & $66.9\pm7.7$ & $0.77\pm0.28$ & $10.81\pm0.04$ \\
858 & 266.64000 & 65.53503 & $21.64\pm0.04$ & $21.33\pm0.07$ & $26.0\pm2.7$ & $0.72\pm0.25$ & $11.31\pm0.01$ \\
870 & 266.69924 & 65.53562 & $22.25\pm0.08$ & $21.94\pm0.11$ & $14.4\pm3.9$ & $0.30\pm0.28$ & $10.93\pm0.03$ \\
874 & 266.90847 & 65.53584 & $22.39\pm0.09$ & $21.83\pm0.10$ & $55.0\pm5.3$ & $0.91\pm0.28$ & $10.63\pm0.03$ \\
930 & 266.79381 & 65.53762 & $22.77\pm0.12$ & $22.19\pm0.14$ & $41.8\pm6.9$ & $0.47\pm0.26$ & $10.32\pm0.06$ \\
1002 & 266.68019 & 65.54095 & $22.40\pm0.09$ & $21.62\pm0.08$ & $83.6\pm6.2$ & $1.11\pm0.25$ & $10.73\pm0.03$ \\
1006 & 266.82164 & 65.54123 & $22.92\pm0.14$ & $22.31\pm0.15$ & $70.6\pm10.7$ & $0.49\pm0.23$ & $10.45\pm0.06$ \\
1043\tablenotemark{\dag} & 266.80809 & 65.54326 & $18.73\pm0.00$ & $17.35\pm0.00$ & $298.4\pm0.5$ & $108.00\pm0.28$ & $12.33\pm0.00$ \\
1046 & 266.81166 & 65.54294 & $21.73\pm0.05$ & $21.51\pm0.08$ & $20.7\pm2.6$ & $0.72\pm0.30$ & $11.38\pm0.01$ \\
1085 & 266.77394 & 65.54452 & $22.84\pm0.13$ & $22.00\pm0.12$ & $192.7\pm16.4$ & $1.49\pm0.29$ & $10.30\pm0.07$ \\
1151 & 266.80895 & 65.54695 & $22.21\pm0.07$ & $21.59\pm0.08$ & $73.8\pm5.7$ & $0.98\pm0.24$ & $10.78\pm0.03$ \\
1187 & 266.77843 & 65.54844 & $22.51\pm0.10$ & $22.12\pm0.13$ & $31.7\pm6.4$ & $0.34\pm0.23$ & $10.69\pm0.03$ \\
1193 & 266.85794 & 65.54868 & $22.54\pm0.10$ & $21.82\pm0.10$ & $82.5\pm7.6$ & $0.85\pm0.24$ & $10.21\pm0.06$ \\
1247 & 266.90870 & 65.55051 & $22.85\pm0.13$ & $22.01\pm0.12$ & $87.3\pm9.0$ & $0.83\pm0.26$ & $10.10\pm0.07$ \\
1248 & 266.81848 & 65.55068 & $21.61\pm0.04$ & $21.19\pm0.06$ & $49.8\pm2.6$ & $1.77\pm0.30$ & $11.28\pm0.01$ \\
1249 & 266.82366 & 65.55023 & $22.43\pm0.09$ & $21.70\pm0.09$ & $64.2\pm4.5$ & $1.43\pm0.32$ & $10.87\pm0.03$ \\
1259 & 266.91055 & 65.55101 & $21.53\pm0.04$ & $21.10\pm0.05$ & $30.2\pm2.2$ & $1.13\pm0.28$ & $11.32\pm0.01$ \\
1263 & 266.63926 & 65.55091 & $23.88\pm0.31$ & $22.32\pm0.15$ & $309.2\pm59.5$ & $0.75\pm0.23$ & $9.93\pm0.15$ \\
1292 & 266.87637 & 65.55294 & $21.52\pm0.04$ & $21.09\pm0.05$ & $50.8\pm1.9$ & $3.26\pm0.39$ & $11.41\pm0.01$ \\
1304 & 266.78208 & 65.55290 & $22.95\pm0.14$ & $22.19\pm0.14$ & $101.7\pm12.0$ & $0.71\pm0.24$ & $10.23\pm0.07$ \\
1314 & 266.78424 & 65.55358 & $23.29\pm0.19$ & $22.55\pm0.19$ & $170.4\pm25.7$ & $0.64\pm0.24$ & $10.07\pm0.09$ \\
1338 & 266.84831 & 65.55210 & $23.23\pm0.18$ & $22.56\pm0.19$ & $382.4\pm53.3$ & $1.31\pm0.29$ & $9.95\pm0.10$ \\
1339 & 266.84808 & 65.55223 & $23.61\pm0.25$ & $22.51\pm0.18$ & $796.6\pm221.8$ & $1.07\pm0.25$ & $10.10\pm0.09$ \\
1369 & 266.81018 & 65.55505 & $23.19\pm0.18$ & $22.55\pm0.19$ & $73.6\pm13.8$ & $0.36\pm0.21$ & $9.93\pm0.09$ \\
1400 & 266.71106 & 65.55620 & $23.07\pm0.16$ & $22.22\pm0.14$ & $119.2\pm13.6$ & $0.77\pm0.24$ & $10.41\pm0.06$ \\
1412 & 266.80336 & 65.55674 & $22.79\pm0.12$ & $22.20\pm0.14$ & $58.2\pm8.2$ & $0.54\pm0.25$ & $10.70\pm0.04$ \\
1456 & 266.72274 & 65.55935 & $22.89\pm0.14$ & $22.09\pm0.13$ & $98.3\pm11.0$ & $0.70\pm0.23$ & $10.43\pm0.05$ \\
1461 & 266.90942 & 65.55982 & $22.52\pm0.10$ & $21.59\pm0.08$ & $102.4\pm5.8$ & $2.00\pm0.33$ & $10.85\pm0.03$ \\
1502 & 266.87502 & 65.56179 & $22.82\pm0.13$ & $22.04\pm0.12$ & $158.6\pm14.9$ & $1.12\pm0.26$ & $10.31\pm0.07$ \\
1509 & 266.73619 & 65.56193 & $22.67\pm0.11$ & $21.55\pm0.08$ & $124.0\pm7.9$ & $1.66\pm0.28$ & $10.46\pm0.04$ \\
1521 & 266.64455 & 65.56226 & $22.25\pm0.08$ & $21.19\pm0.06$ & $144.1\pm6.9$ & $2.00\pm0.25$ & $10.87\pm0.02$ \\
1554 & 266.68226 & 65.56372 & $22.34\pm0.08$ & $21.38\pm0.07$ & $178.8\pm9.0$ & $2.37\pm0.28$ & $10.66\pm0.03$ \\
1555 & 266.68152 & 65.56408 & $22.85\pm0.13$ & $21.77\pm0.10$ & $165.7\pm11.3$ & $1.94\pm0.31$ & $10.87\pm0.04$ \\
1567 & 266.68369 & 65.56430 & $21.97\pm0.06$ & $21.16\pm0.06$ & $94.7\pm4.0$ & $2.41\pm0.30$ & $11.00\pm0.02$ \\
1576 & 266.69107 & 65.56451 & $22.77\pm0.12$ & $22.14\pm0.13$ & $116.1\pm11.6$ & $0.90\pm0.25$ & $10.48\pm0.05$ \\
1887 & 266.76772 & 65.57359 & $22.57\pm0.10$ & $22.06\pm0.12$ & $56.2\pm6.2$ & $0.82\pm0.29$ & $10.38\pm0.05$ \\
2126 & 266.69413 & 65.50960 & $22.81\pm0.13$ & $21.91\pm0.11$ & $86.8\pm7.5$ & $1.11\pm0.29$ & $10.47\pm0.04$ \\
\hline
\end{tabular}
\vspace{3mm}
\tablecomments{$^{\dag}$The ID\#1043 is the radio galaxy 4C\,65.22 at $z=1.52$ (i.e.\ the primary target of this study). $^{\ddag}$The $M_{\star}$ uncertainties provided here are based solely on the photometric uncertainties in $K_s$-band. We note, however, that the uncertainties in $M_{\star}$ could potentially be dominated by the systematic uncertainties associated with the derivation of $M_{\star}$ using the one-color method (typically $\pm$0.2~dex; see \citealt{koy13b}).}
\end{center}
\end{table*}

\subsection{Photometric redshifts and the final member catalog}

\begin{figure}
 \begin{center}
 \vspace{0.3cm}
 \rotatebox{270}{\includegraphics[height=7.8cm]{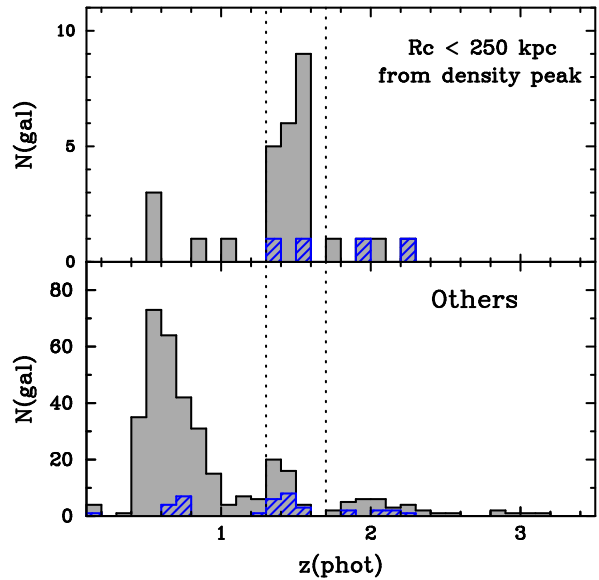}}
 \end{center}
 \vspace{-0.3cm}
\caption{ The $z_{\rm phot}$ distribution of galaxies for those with $z_{\rm phot}$$<$3.5 located within 250~kpc from the density peak (top) and those outside this annulus (bottom), showing a clear redshift spike at $z\sim 1.5$ in the 4C65.22 field. The dotted lines at $z_{\rm phot}=$1.3 and 1.7 show the criteria of photo-$z$ member selection applied in this paper. The blue hatched histogram shows $z_{\rm phot}$ distribution for the H$\alpha$ emitters selected in \S~3.1. The photo-$z$ accuracy for SF galaxies should be less reliable, but we emphasize that the primary aim of photo-$z$ is to select passive galaxies without H$\alpha$ emission. \\
\label{fig:zphot_histogram}}
\end{figure}

\begin{figure*}
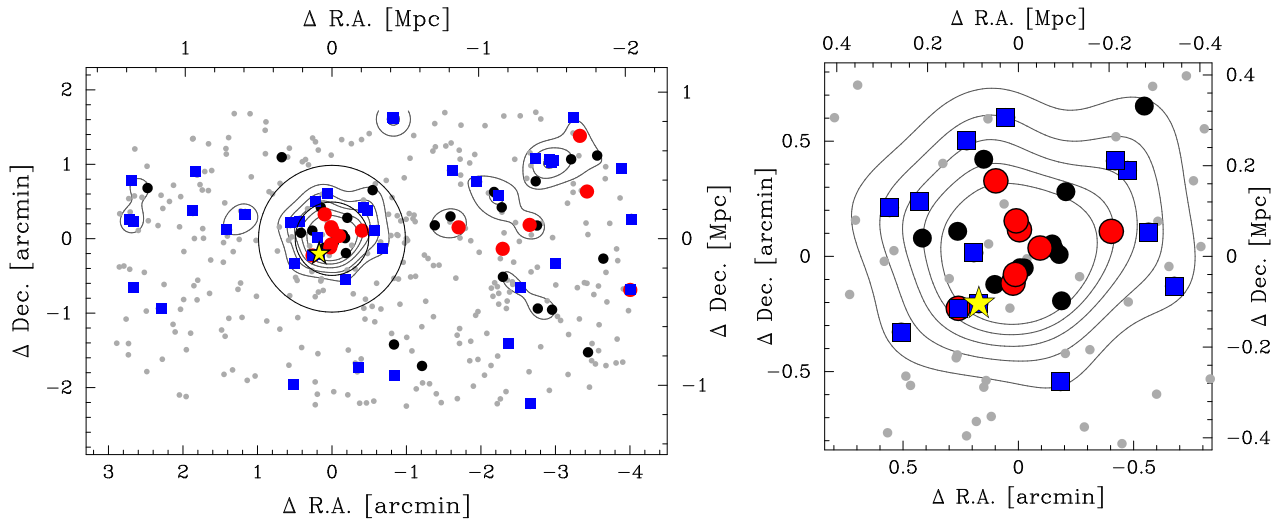

 \begin{center}
 \vspace{0.0cm}
 \includegraphics[angle=270.0,scale=0.52]{fig4a.ps}
 \hspace{1mm}
 \includegraphics[angle=270.0,scale=0.52]{fig4b.ps}
 \end{center}
 \vspace{-0.2cm}
\caption{ The 2-D map of galaxies around 4C\,65.22 over the entire MOIRCS FoV (4$'\times$7$'$; left panel) and the 100$''\times$100$''$ close-up view around the density peak (right panel). The grey dots, black circles, and blue squares show all NB-detected sources, photo-$z$ selected galaxies with $1.3\le z_{\rm phot} \le 1.7$, and the H$\alpha$ emitters at $z=1.5$, respectively. The red symbols show the galaxies on the red sequence. The red-sequence galaxies are strongly clustered near the radio galaxy (yellow star), and H$\alpha$ emitters tend to be located in the cluster outskirts. The contours show 1,2,3,4,5$\sigma$ significance of the overdensity calculated with all member galaxies: we applied Gaussian smoothing ($\sigma$=0.1~Mpc) on each galaxy and combine the tails of Gaussian wings to measure local density at a given point. The (0,0) point of this plot shows the density peak (RA$=$17$^h$47$^m$12.3$^s$, Dec$=$+65$^d$32$^m$48$^s$). The solid-line circles in the left-hand panel show 250 and 500~kpc from the density peak. \\ 
\label{fig:map}}
\end{figure*}

We use photometric redshifts ($z_{\rm phot}$) to select non-H$\alpha$-emitting passive members (with strong 4000\AA\ break) which cannot be identified with the above H$\alpha$ emitter selection. We here restrict the sample to those detected at $>$2$\sigma$ level with at least four bands out of the six broad-band images ($Br'z'JHK_s$). The best-fit photometric redshifts are computed with a simple template fitting following the standard $\chi^2$ minimizing statistics using EAZY code (\citealt{bra08}). The distribution of the derived $z_{\rm phot}$ is shown in Fig.~3. The photometric redshifts derived with only six broad-band filters may not be very accurate, but it is worth mentioning that there is a clear redshift spike at $z_{\rm phot}$$\sim$1.5 near the radio galaxy (top panel of Fig.\,\ref{fig:zphot_histogram}). In the following analysis, we apply a photo-$z$ cut of $1.3<z_{\rm phot}<1.7$ (dotted lines in Fig.\,\ref{fig:zphot_histogram}) to select member galaxies associated to the radio galaxy. With this criteria, we select 61 galaxies in total, among which 20 galaxies are within the central 250~kpc region\footnote{We compute the density peak using all the member galaxies, including photo-$z$ members and H$\alpha$ emitters, and exploit this point as the cluster center (see also Fig.~\ref{fig:map}).}. 

As shown in Fig.~\ref{fig:color_select} (black circles), most of the photo-$z$ selected galaxies satisfy the {\it BzK} and {\it rJK} criteria. Unfortunately, we cannot fully quantify the photo-$z$ accuracy because no spec-$z$ information is available in the observed field. However, we emphasize that our results do not change even if we apply simple color selections without relying on the photometric redshift. We here crudely test the contamination level by using the UKIDSS/UDS data, where $BRzJHK$ data are publicly available (\citealt{law07})\footnote{http://www.nottingham.ac.uk/astronomy/UDS/index.html}. After applying the same magnitude cuts, we derive photo-$z$ of the UDS galaxies in the same way as described above. We find that $\sim$2250 galaxies (out of $\sim$25000 sources) satisfy the $1.3<z_{\rm phot}<1.7$ criteria over the entire $\sim$0.8~deg$^2$ UDS field, yielding the expected contamination rate of $\sim$0.8 arcmin$^{-2}$. Based on this estimate, we expect $\sim$0.6 galaxies in the central 250 kpc region. Therefore, the central 250 kpc region of 4C65.22 field is indeed over-dense by a factor of $\sim$30$\times$ compared to the control field (see also \S~4). We also expect $\sim$22 galaxies within 4$'\times$7$'$ MOIRCS FoV. The number density of the photo-$z$ members in the 4C65.22 field is still significantly higher than general field (by a factor of $\sim$3$\times$), but this analysis suggests that the over-density is diluted when averaged over the FoV.   

For star-forming galaxies, the photometric redshifts may be much less reliable because of their flat featureless broad-band SEDs. We find that only 20 H$\alpha$ emitters (45\%) satisfy the photo-$z$ criteria of 1.3$<$$z_{\rm phot}$$<$1.7 (see blue-hatched histograms in Fig.\,\ref{fig:zphot_histogram}). In particular, we find that the photometric redshift does not work well for H$\alpha$ emitters with red optical colors, and that the photo-$z$ errors of such red emitters tend to be slightly larger than those of blue emitters. This probably reflects the well-known age/dust degeneracy, making it extremely difficult to construct a complete sample of SF galaxies using a limited set of broad-band filters. We remind the readers that this point is an important advantage of the narrow-band survey: it allows us to construct a clean star-forming galaxy sample within a narrow redshift slice at a high completeness level (e.g.\ \citealt{kod04}; \citealt{koy10}). In the case of our current dataset, $\sim$55\% of SF galaxies might have been missed out with the photo-$z$ selection alone, but those galaxies are now {\it rescued} with our NB-based approach. 

In summary, we selected the final member galaxies with the following two criteria. First, we select 61 galaxies with $1.3< z_{\rm phot} < 1.7$. Second, we include all 44 H$\alpha$ emitters (selected in \S~3.1) regardless of their photometric redshifts. Note that 20 out of the 44 H$\alpha$ emitters satisfy the above photo-$z$ criteria as well. In these ways, our final member catalog contains 85 galaxies in total, and this sample will be used in the remaining part of this paper. We note that a fraction of the selected H$\alpha$ emitters might be AGNs. In general, the fraction of AGNs in H$\alpha$-selected galaxies at $z\gsim 1$ is expected to be $\sim$10--15\% (\citealt{gar10}; \citealt{sob13}). However, unfortunatately, it is not possible to quantify the AGN contribution in our own sample, without multi-wavelength imaging/spectroscopic follow-up observations. In particular, the environmental dependence of the AGN activity in the distant universe is still an open question and beyond the scope of this paper. 

\begin{figure}
 \begin{center}
 \vspace{0.2cm}
 \includegraphics[angle=0.0,scale=0.22]{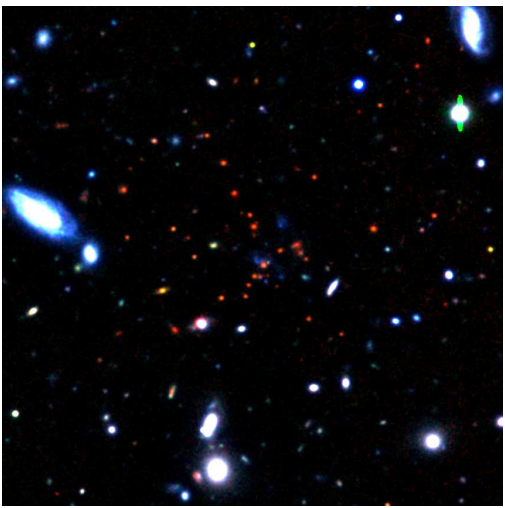}
 \end{center}
 \vspace{-0.2cm}
\caption{ The false-color image of the 4C65.22 field created with $r'z'H$-band data. The size of this image is 100$''\times$100$''$, same as the right-hand panel of Fig.~4. North is up, and east is to the left. \\ 
\label{fig:image}}
\end{figure}


\section{Discovery of a rich cluster at z=1.52}

\subsection{Spatial distribution of galaxies around 4C65.22}

Based on the sample selection described in \S~3, we here report the discovery of a strong over-density of $z\sim 1.5$ galaxies around the radio galaxy 4C\,65.22. In Fig.~\ref{fig:map}, we show the spatial distribution of the selected member galaxies. We here plot photo-$z$ selected galaxies (with 1.3$<$$z_{\rm phot}$$<$1.7; black circles) and the H$\alpha$ emitters (blue squares). The member galaxies on the red sequence ($z'-J>1.45$; see \S~4.2) are marked with red symbols. It is evident that there is a clear density peak of $z\sim 1.5$ galaxies near the radio galaxy. The radio galaxy (shown with yellow star in Fig.~\ref{fig:map}) is located at $\approx$140 kpc away from the density peak, not in the very center of the core. 

We already showed in Fig.~\ref{fig:zphot_histogram} a clear redshift spike at $z_{\rm phot}\sim 1.5$ within $r_c$$<$250~kpc from the density peak. This result does not change even if we consider ``all'' member galaxies including H$\alpha$ emitters. The number density of all the member galaxies (i.e. photo-$z$ member and H$\alpha$ emitters) within the 250~kpc circle is 28.9~arcmin$^{-1}$, which is a factor of $>$10$\times$ higher than the average in the outer field (2.3 arcmin$^{-2}$). We note that a small $z_{\rm phot}$ peak at $z\sim 1.5$ can also be seen in the bottom panel of Fig.~\ref{fig:zphot_histogram}. This probably reflects the presence of large-scale structures extending beyond the 250~kpc circle. Although we need spectroscopic follow-up observation for confirming such relatively poor structures, it is highly possible that some of the small groups as well as the filamentary structures traced by H$\alpha$ emitters seen in Fig.~\ref{fig:map} are really associated to the central cluster.

\subsection{Red-sequence galaxies dominating the cluster core \\ 
and the deficit of low-mass blue galaxies}

Another important result drawn from Fig.~\ref{fig:map} is that the cluster central region ($r_c$$\lsim$250 kpc) is clearly dominated by red-sequence galaxies without H$\alpha$ emission (see also Fig.~\ref{fig:image} for a three-color representation of our $rzH$ images in the central cluster region). Interestingly, this ``suppression radius'' of $\sim$200~kpc is consistent with that reported in another X-ray selected rich cluster at a similar redshift (e.g.\ \citealt{bau11a}), suggesting this newly discovered system is a well-evolved rich galaxy cluster at $z=1.5$. The presence of a number of H$\alpha$ emitters (at $1.51\le z \le1.54$) surrounding the core further supports our discovery of the rich cluster associated to the radio galaxy. We note that no extended X-ray emission is reported in this field so far; e.g. by the ROSAT all-sky cluster survey (\citealt{boh00}). This is not surprising given the shallowness of the all-sky survey data. However, it would be highly possible that we can detect an extended X-ray emission associated with this cluster if we can perform a deep X-ray follow-up observation in the future.

\begin{figure}
 \begin{center}
 \vspace{0.3cm}
 \rotatebox{270}{\includegraphics[height=7.8cm]{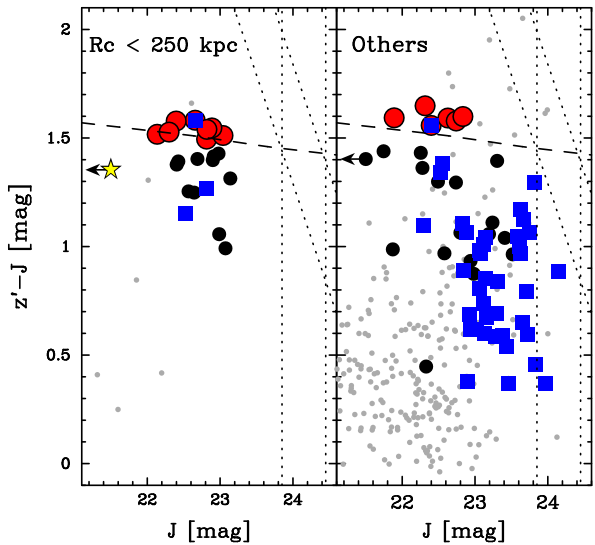}}
 \end{center}
 \vspace{-0.2cm}
\caption{ The color--magnitude diagram for all NB-selected sources within 250~kpc from the density peak (left) and outside this radii (right). Black circles show the member galaxies, while blue squares show H$\alpha$ emitters. The red circles are member galaxies with $z'-J>1.45$ (identical to the red symbols in Fig.~4). The vertical and slanted dotted lines show 3- and 5-$\sigma$ limiting magnitudes in $z'$ and $J$, respectively. The dashed line shows the location of the red sequence (assuming $z_f=5$) modelled by Kodama et al. (1998). Interestingly, low-mass blue galaxies are clearly deficient in the cluster core region. \\
\label{fig:colmag}}
\end{figure}

In Fig.~\ref{fig:colmag}, we show color--magnitude diagrams for galaxies located at $r_c$$<$250~kpc and those outside this annulus. An interesting feature recognized in this plot is the clear ``deficit'' of blue galaxies in the central cluster region ($r_c$$<$250~kpc), in stark contrast to the outer field. There still exist a few H$\alpha$ emitters within the $r_c=$250~kpc circle, but these H$\alpha$ emitting galaxies in the cluster core region show redder colors than typical H$\alpha$ emitters in the field environment. This clear lack of low-mass blue galaxies may represent an accelerated galaxy evolution in high-$z$ cluster environments; it is most likely explained that blue SF galaxies falling into the cluster are quickly transformed into red/massive quiescent population. It is also likely that the quenching process accompanies a quick mass growth via e.g. mergers or galaxy--galaxy interactions, otherwise we cannot fully explain the deficit of low-mass blue galaxies {\it and} the excess of red massive galaxies in the cluster core region at the same time. In Fig.~\ref{fig:colmag}, the red sequence is only visible in its bright end, consistent with some recent studies of $z\sim 1$ clusters (e.g. \citealt{koy07}; \citealt{sto07}). The lack of the faint end of the red sequence is unlikely due to the selection effect: we verified that the number of faint red galaxies does not increase even if we use $J$-band detected catalog instead of NB-selected catalog. 

\begin{figure}
 \begin{center}
 \vspace{0.3cm}
 \rotatebox{270}{\includegraphics[height=7.8cm]{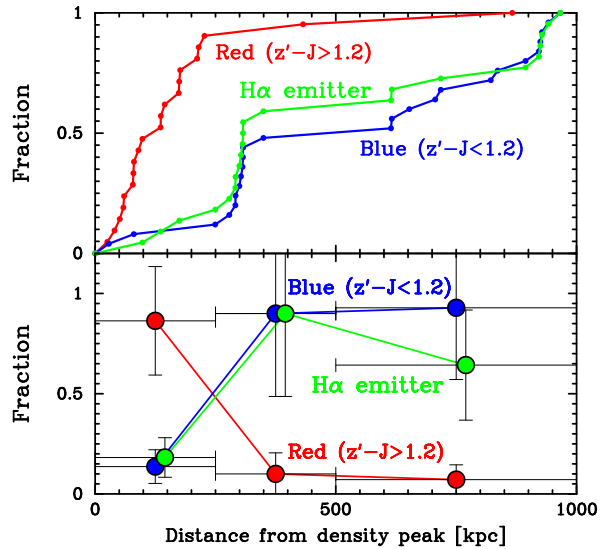}}
 \end{center}
 \vspace{-0.2cm}
\caption{ The cumulative (top) and differential (bottom) plot of the fraction of each population to the total galaxy sample studied here as a function of the distance from the density peak. This plot clearly demonstrates that the dominant galaxy population sharly changes at around $r_c$$\sim$200--300~kpc. In the bottom panel, the vertical error-bars represent 1$\sigma$ uncetrainty from Poisson statistics, while the horizontal error-bars show the bin size used for calculation. We applied a small shift for the symbols for clarity. Most blue galaxies and H$\alpha$ emitters are overlapped with each other, but they are not completely overlapped: i.e. there are some red H$\alpha$ emitters as well as blue galaxies without H$\alpha$ emission. \\
\label{fig:fraction}}
\end{figure}

We also show in Fig.~\ref{fig:fraction} the cumulative (and differential) fraction of each population in our total sample as a function of the distance from the density peak out to $r_c=$1~Mpc. This plot further supports our claim that red galaxies are strongly clustered within $r_c\sim$250~kpc. It is evident that the fraction of red galaxies sharply drops at $r_c\sim 200-300$kpc, which coincides with the place where we see the dramatical increase in H$\alpha$ emitters and blue galaxies. The Kolmogorov-Smirnov test (KS-test) suggests with $>$99.9\% confidence level that the distributions of red galaxies and H$\alpha$ emitters (or blue galaxies) are drawn from a different parent population. We therefore conclude that most of blue star-forming galaxies falling into the highest-density cluster core are forced to migrate toward the red sequence, resulting in the clear deficit of low-mass blue galaxies in the central part of this cluster. 

\section{Galaxy properties v.s. Environment}

\subsection{Color--density and M$_{\star}$--density relation}

Using the newly discovered cluster field as a laboratory, we here take a more detailed look at the environmental dependence of galaxy properties at $z\sim 1.5$. For each member galaxy, we calculate the local surface number density ($\Sigma_{\rm 5th}=5/\pi r_{\rm 5th}^2$), where $r_{\rm 5th}$ denotes the distance to the fifth-nearest neighbor of each member galaxy. In the following analysis of this paper, we use this local density $\Sigma_{\rm 5th}$ as an environmental index, but our results do not change even if we use $\Sigma_{\rm 3rd}$ or $\Sigma_{\rm 10th}$.

\begin{figure*}
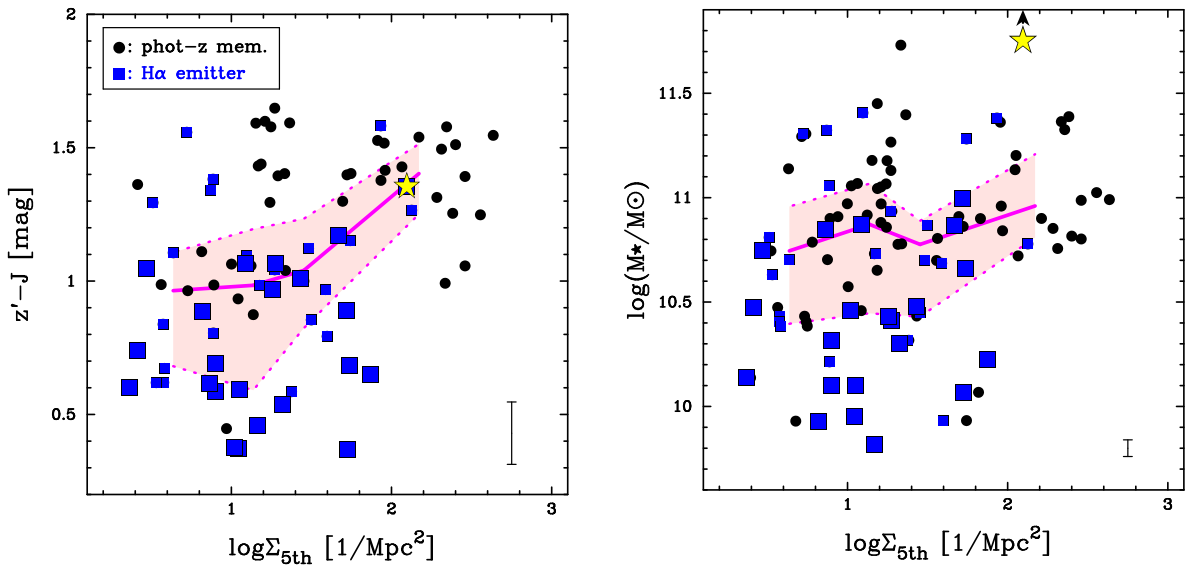

 \begin{center}
 \vspace{0.3cm}
 \includegraphics[angle=270,scale=0.43]{fig8a.ps}
 \hspace{5mm}
 \includegraphics[angle=270,scale=0.43]{fig8b.ps}
 \end{center}
 \vspace{-0.3cm}
\caption{ The color--density and $M_{\star}$--density relation for $z\sim 1.5$ galaxies around the radio galaxy 4C\,65.22. The blue squares and black circles show the selected member galaxies with and without H$\alpha$ emission, respectively. We split the H$\alpha$ emitter sample into two equal sized bins based on EW$_{\rm rest}$(H$\alpha$+[N{\sc ii}]), and use large and small squares for high-EW and low-EW emitters, respectively. We also show the running median in each plot calculated by splitting the sample into four equal-sized bins (magenta solid line), as well as the 25\% and 75\% distribution by the pink shaded region. These plots show that the color and $M_{\star}$ are both correlated with the local galaxy density, but we note that the environmental trends become less prominent when we consider only H$\alpha$ emitting galaxies. Typical uncertainties are indicated at the bottom-right corner of each panel. The uncertainty in $M_{\star}$ shown here simply reflects the $K_s$-band photometric errors (hence lower limits), but we note that the systematic uncertainties accompanied by the derivation of $M_{\star}$ with the one-color method (i.e.\ Eq. 1) may potentially dominate the $M_{\star}$ uncertainty (typically $\pm$0.2~dex as discussed by \citealt{koy13b}).\\
\\
\label{fig:color_density}}
\end{figure*}

We first focus on galaxy colors. In Fig.\,\ref{fig:color_density} (left), we show $z'-J$ (rest-frame $U-V$) colors of member galaxies as a function of $\Sigma_{\rm 5th}$. This plot clearly demonstrates that the color--density relation is already in place at $z\sim 1.5$, consistent with the visual impression from Fig.~\ref{fig:map}. We here compute the Spearman's rank correlation coefficient ($\rho=0.40$). With a sample size of $N_{\rm all}$$=$84, we can conclude that there is a significant correlation between the two variables\footnote{In the following statistical tests, we do not use the radio galaxy itself because its physical quantities such as color, mass, and SFR are contaminated by unpredictable central AGN activity. However, our results do not change even if we include it in the analyses.}: a null hypothesis that there is no correlation between $z'-J$ and $\Sigma_{\rm 5th}$ is ruled out at $>$99\% confidence level.

The correlation becomes much less significant when we consider only H$\alpha$ emitters. In this case, the Spearman's rank correlation coefficient is $\rho=0.11$. With a sample size of $N_{\rm HAE}=43$, we cannot rule out a null hypothesis that there is no significant correlation between $z'-J$ and $\Sigma_{\rm 5th}$ for H$\alpha$ emitters. Therefore we suggest that the color--density relation is primarily driven by passive red galaxies in the highest-density cluster core, equivalently by the lack of blue galaxies in the highest-density region. We note, however, that there is a weak correlation between the EW(H$\alpha$+[N{\sc ii}]) and galaxy colors in the sense that redder H$\alpha$ emitters tend to have lower EW (see blue squares in Fig.~8, where we use different symbol size for high-EW/low-EW emitters). In other words, our definition of the H$\alpha$ emitters ($H-$NB1657$>$0.2) may potentially fail to select low-EW (red) H$\alpha$ emitters. Therefore we need to keep in mind that the results on the color--density relation amongst H$\alpha$ emitters could be highly sensitive to the EW cut used for the sample selection.


It is possible that the color--density relation is partly produced by the correlation between galaxy stellar mass and environment. We here investigate the environmental dependence of stellar masses ($M_{\star}$) of the member galaxies. The $M_{\star}$ of galaxies are derived using $K_s$-band photometry with $M_{\star}/L_{\rm K_s,obs}$ correction based on the $z'-K_s$ color as the following. 
\begin{equation}
\log (M_{\star}/10^{11}M_{\odot})_{z=1.5} = -0.4(K_s-21.53) + \Delta M_{\star},
\end{equation}
where $\Delta M_{\star} = 0.095 - 1.003\times\exp(-0.807\times(z'-K_s))$ indicates the $M/L$ correction value. These conversion equations are derived using the model galaxies developed by \cite{kod98} assuming \cite{sal55} IMF. It is demonstrated that this ``one-color method'' can provide reasonable $M_{\star}$ values out to $z\sim 2$ (see \citealt{koy13b}). 

In Fig.~\ref{fig:color_density} (right), we plot the derived $M_{\star}$ as a function of $\Sigma_{\rm 5th}$. Our data suggest that there exists a weak $M_{\star}$--density relation in the observed field, again driven by the massive galaxies in the highest-density cluster core. By computing the Spearman's rank correlation coefficient ($\rho=0.23$), and with a sample size of $N_{\rm all}=$84, we can rule out the null hypothesis that there is no $M_{\star}$--$\Sigma_{\rm 5th}$ relationship with $\approx$97\% confidence level. However, if we focus only on the H$\alpha$ emitters, the Spearman's rank correlation coefficient is $\rho=0.11$, suggesting no significant correlation between the two variables with the sample size of $N_{\rm HAE}=43$. Our data thus suggest that the color--density and $M_{\star}$--density relation are already in place at $z\sim 1.5$, mostly driven by passive red/massive galaxies in the highest-density region. However, the correlations almost disappear when we consider only SF galaxies.

\subsection{SFR--density and SSFR--density relation}

An important advantage of the NB H$\alpha$ survey is that we can measure H$\alpha$ flux of all emitters within the observed FoV. We here describe how we derive the H$\alpha$-based SFRs of H$\alpha$ emitters, and then describe how the derived SFRs depend on environment. First of all, we calculate the H$\alpha$+\nii\ line flux ($F_{\rm{H}\alpha + \rm{[NII]}}$), continuum flux density ($f_c$) and rest-frame equivalent width (EW$_{\rm rest}$) of each H$\alpha$ emitter based on its broad-band and narrow-band photometry with the following equations:
\begin{equation}
F_{\rm{H\alpha + [NII]}}=\Delta_{\rm{NB}}\frac{ f_{\rm{NB}} -
 f_{H} }{1-\Delta_{\rm{NB}}/\Delta_{H}}
\end{equation}
\begin{equation}
f_c = \frac{f_{H} - f_{\rm{NB}}(\Delta_{\rm{NB}}/\Delta_{H})}{1-\Delta_{\rm{NB}}/\Delta_{H}}
\end{equation}
\begin{equation}
\textrm{EW}_{\rm rest}({\rm{H\alpha + [NII]}}) = (1+z)^{-1} \frac{F_{\rm{H\alpha}
 + [NII]}}{f_c}. 
\end{equation}
The $\Delta_{H}$ ($=0.28$$\mu$m) and $\Delta_{\rm{NB}}$ ($=0.0195$$\mu$m) are the FWHMs of the broad-band ($H$) and NB1657 filters, $f_{H}$ and $f_{\rm{NB}}$ are the flux densities at $H$-band and at NB1657, respectively. We then multiply 4$\pi$$d_L^2$ by $F_{\rm{H\alpha +[NII]}}$ to derive the luminosity $L_{\rm{H\alpha +[NII]}}$, where $d_L$ is the luminosity distance of 1.109$\times$10$^3$ Mpc at $z=1.5$.

\begin{figure*}
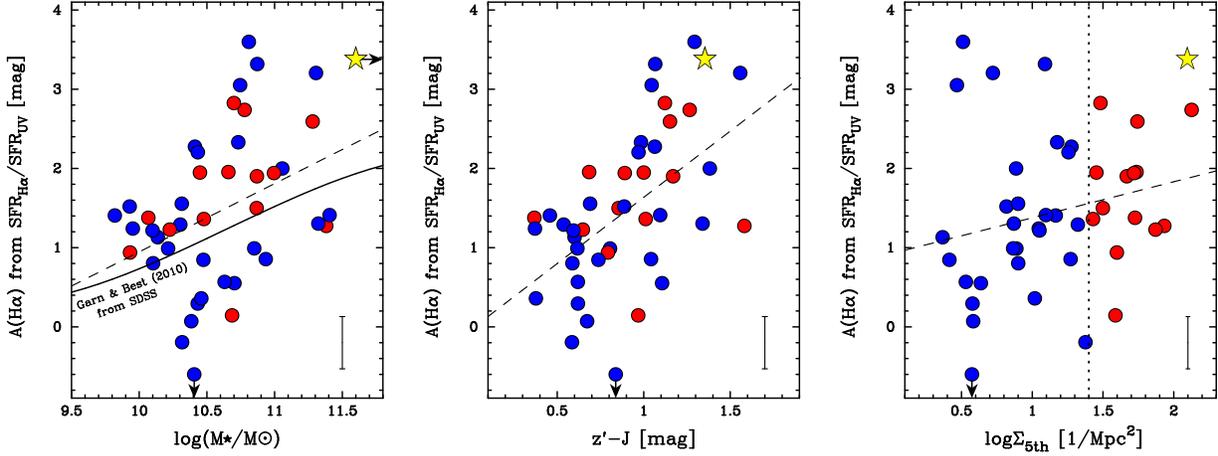

 \begin{center}
 \vspace{0.4cm}
 \includegraphics[angle=270,scale=0.38]{fig9a.ps}
 \hspace{3mm}
 \includegraphics[angle=270,scale=0.38]{fig9b.ps}
 \hspace{3mm}
 \includegraphics[angle=270,scale=0.38]{fig9c.ps}
 \end{center}
 \vspace{-0.3cm}
\caption{ The $A_{\rm H\alpha}$ values derived from SFR(H$\alpha$)/SFR(UV) ratio against their stellar mass (left), $z'-J$ colors (middle), and environment (right). The red and blue symbols show the H$\alpha$ emitters in the high-density ($\log\Sigma_{\rm 5th}\ge$1.4; top $\sim$1/3) and in low-density ($\log\Sigma_{\rm 5th}<$1.4) environment. The yellow star indicates the radio galaxy. In the left-hand panel, we also show the empirical $A_{\rm H\alpha}$--$M_{\star}$ relation for local star-forming galaxies from \cite{garnbest10}, accounted for IMF difference. Our H$\alpha$ emitter sample at $z=1.52$ also shows that the $A_{\rm H\alpha}$ value increases with $M_{\star}$, but there seems to be a large scatter around the $A_{\rm H\alpha}$--$M_{\star}$ relation. In each plot, the dashed line shows the simple best-fitted relation, and the typical error-bar size is shown in the bottom-right part of each plot. Interestingly, there is a weak positive correlation between $A_{\rm H\alpha}$ and $\Sigma_{\rm 5th}$ (although only with $\sim$1$\sigma$ significance), suggesting star-forming galaxies in high-density environment may be more highly obscured by dust. \\
\\
\label{fig:extinction}}
\end{figure*}

We correct for the \nii\ line contribution into the total NB fluxes, using the empirical calibration between \nii/(H$\alpha$+\nii) flux ratio and EW$_{\rm rest}$(H$\alpha$+\nii) for local SDSS galaxies presented by \cite{sob12}. The median of the derived \nii/(H$\alpha$+\nii) ratio for our H$\alpha$ emitter sample is 0.19, consistent with the H$\alpha$ emitter sample at $z=1.46$ (\nii/(H$\alpha$+\nii)$=$0.22) shown by \cite{sob12}. Although there remains a large scatter around this relation ($\sim$0.4~dex; e.g. \citealt{vil08}), it is impossible to measure [N{\sc ii}] line flux for all galaxies because it requires deep spectroscopic observation for all of them. Therefore our procedure is the best effort at this moment, and at least it would be more realistic than the convensional constant (e.g.\ 30\%) correction. We then compute the dust-free H$\alpha$-based star formation rates, SFR$_{\rm H\alpha}$, using the \cite{ken98} relation for \cite{sal55} IMF; SFR$_{\rm H\alpha}$[$M_{\odot}$/yr]$=7.9\times 10^{-42} L_{\rm H\alpha}$[erg\,s$^{-1}$]. We note that our H$\alpha$ emitter selection criteria shown in \S~3.1 approximately correspond to EW$_{\rm{rest}}$(H$\alpha$+\nii)$\gsim$20\AA\ and SFR$_{\rm H\alpha}$$\gsim$3$M_{\odot}$/yr. 

Finally, we correct for the dust extinction effect. We here consider two independent methods to estimate dust extinction at H$\alpha$ ($A_{\rm H\alpha}$). The first method is to use $A_{\rm H\alpha}$--$M_{\star}$ relation established for local galaxies by \cite{garnbest10}. The $A_{\rm H\alpha}$--$M_{\star}$ correlation is likely to be unchanged up to $z\sim 1.5$ (\citealt{sob12}; \citealt{iba13}), and this method is often used in studies of distant galaxies as a convenient method to predict dust extinction effect for individual galaxies.

The second approach is to use SFR$_{\rm H\alpha}$/SFR$_{\rm UV}$ ratio for predicting $A_{\rm H\alpha}$. Because H$\alpha$ is less sensitive to dust extinction effect than rest-frame UV light, the SFR$_{\rm H\alpha}$/SFR$_{\rm UV}$ ratio can be used as an indicator of dust attenuation (\citealt{bua02}; \citealt{tad13}). In our case, the $B$-band flux density ($\lambda_{\rm rest}$$\approx$1800\AA) can be converted to the rest-frame UV luminosity density, and then we derive SFR$_{\rm UV}$ using \cite{ken98} calibration; SFR$_{\rm UV}$$=$1.4$\times$10$^{-28}L_{\nu}$[erg\,s$^{-1}$\,Hz$^{-1}$]. The observed SFR$_{\rm H\alpha}$/SFR$_{\rm UV}$ ratio is translated to the dust extinction value using the following equation:
\begin{equation}
E_{\rm star}(B-V) = \frac{2.5\times \log({\rm SFR}_{\rm H\alpha}/{\rm SFR}_{\rm UV})}{k_{\rm UV}-k_{\rm H\alpha}/f},
\end{equation}
\begin{equation}
A_{\rm H\alpha} = k_{\rm H\alpha} \times E_{\rm star}(B-V) / f.
\end{equation}
We here adopt $k_{\rm H\alpha}$$=$3.33 and $k_{\rm UV}$$=$9.37 from \cite{cal00} law with $R_V$$=$4.05. One thing we need to treat carefully is an extra extinction toward the nebular regions ($f$ value in the above equations). In the \cite{cal00} prescription, the $f$ value was shown to be 0.44 (i.e.\ $A_{\rm H\alpha}=A_{\rm cont}/0.44$). However, recent studies claim that this $f$ value is likely to be different for high-$z$ SF galaxies (e.g.\ \citealt{pri13}). \cite{kas13} empirically linked the SFR$_{\rm H\alpha}$/SFR$_{\rm UV}$ ratio to $E(B-V)$ for $z\sim 1.3$ SF galaxies. They derived the best-fitted value of $f=0.69$, and we adopt this value to our analysis. We note that the method may not be perfect enough (e.g. SFR$_{\rm H\alpha}$/SFR$_{\rm UV}$ flux ratio could depend on galactic age; see \cite{wuy13} for a detailed discussion), but we expect this method is more realistic than the simple $M_{\star}$-dependent correction shown above. 

\begin{figure*}
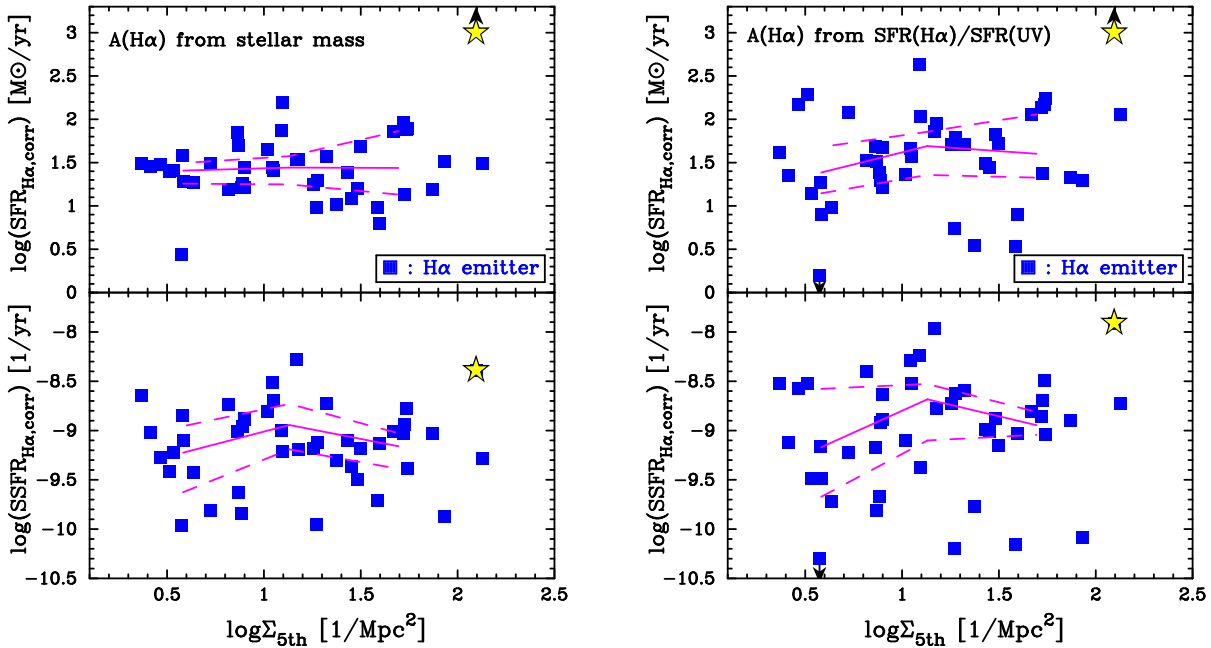

 \begin{center}
 \vspace{0.2cm}
 \includegraphics[angle=270,scale=0.48]{fig10a.ps}
 \hspace{8mm}
 \includegraphics[angle=270,scale=0.48]{fig10b.ps}
 \end{center}
 \vspace{-0.3cm}
\caption{ The SFR--density (top) and SSFR--density (bottom) relation for H$\alpha$ emitters at $z=1.5$. The left and right panel shows the result based on the different dust extinction procedure. Only H$\alpha$ emitters are plotted on these diagrams, because it is impossible to measure SFRs of non-H$\alpha$ emitters. The yellow star indicates the radio galaxy, which will be excluded from the statistical test performed in this paper. Regardless of the dust extinction correction procedure, there is no clear environmental dependence of SFR or SSFR when we consider only star-forming galaxies. The running median (as well as the 25\% and 75\% distribution) are shown in each plot. \\ 
\label{fig:SFR_density}}
\end{figure*}
\begin{figure*}
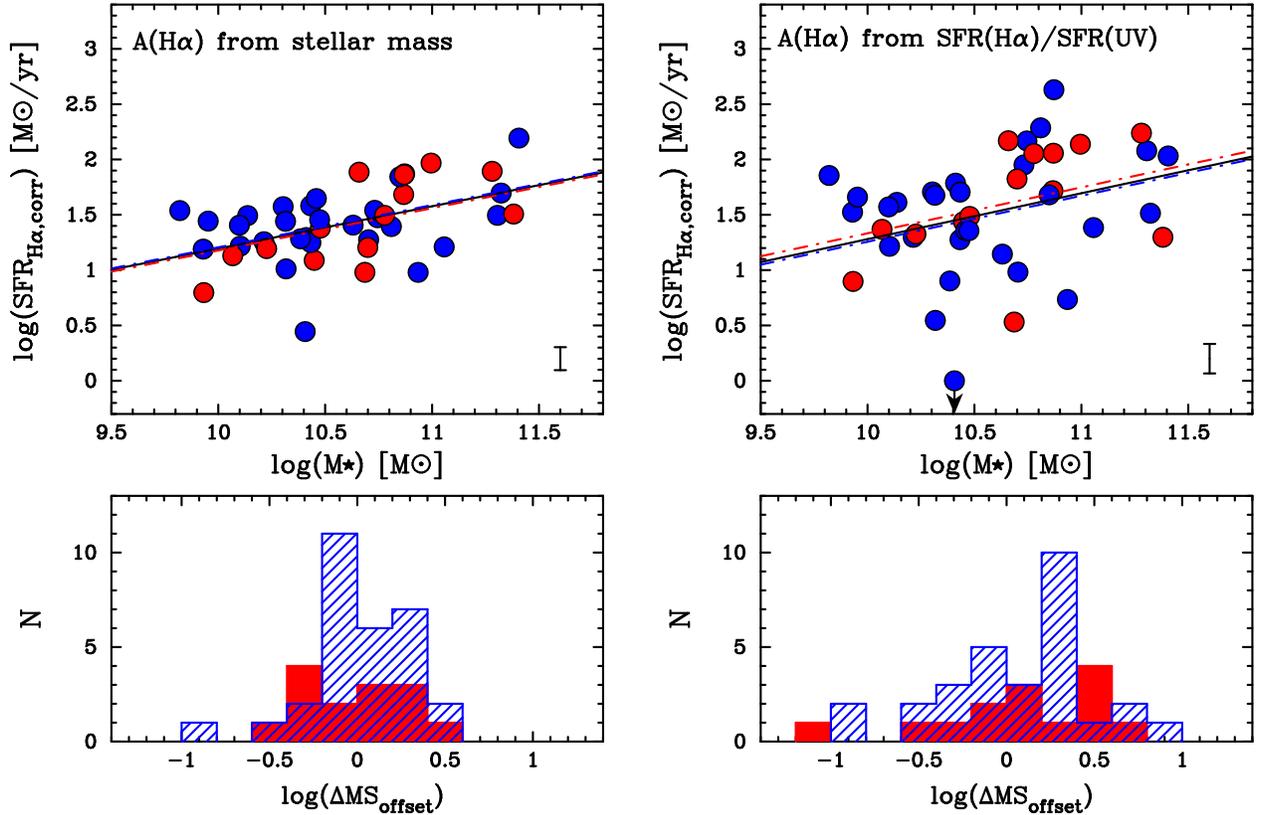

 \begin{center}
 \vspace{0.5cm}
 \includegraphics[angle=270,scale=0.55]{fig11a.ps}
 \hspace{5mm}
 \includegraphics[angle=270,scale=0.55]{fig11b.ps}
 \end{center}
 \vspace{-0.4cm}
\caption{ The SFR--$M_{\star}$ diagram for H$\alpha$ emitters at $z\sim 1.5$ in the field of 4C\,65.22. The left-hand panel shows the result with the $M_{\star}$-dependent extinction correction shown by \cite{garnbest10}, while the right-hand panel shows the result with extinction correction from SFR(H$\alpha$)/SFR(UV) ratio. The radio galaxy is not plotted on this diagram given the large uncertainty for deriving both $M_{\star}$ and SFR. For each plot, we show the best-fitted SFR--$M_{\star}$ relation for all emitters (black solid line), for those in high-density environment ($\log\Sigma_{\rm 5th}\ge 1.4$; red dot-dashed line) and low-density environment ($\log\Sigma_{\rm 5th}< 1.4$; blue dot-dashed line). In the bottom, we show the histograms of the SFR offset from the black line in the top panels ($\Delta$MS$_{\rm offset}$) for each environmental bin; red filled histogram for high-density and blue shaded histogram for low-density environment. \\
\label{fig:MS}}
\end{figure*}

In Fig.~\ref{fig:extinction} (left), we plot the derived $A_{\rm H\alpha}$ values against their stellar mass for all H$\alpha$ emitters. The extinction effect (derived by the SFR$_{\rm H\alpha}$/SFR$_{\rm UV}$ ratio) increases with stellar mass, showing a reasonable agreement with the local $A_{\rm H\alpha}$--$M_{\star}$ relation. Also, the derived $A_{\rm H\alpha}$ values are higher for red H$\alpha$ emitters (see middle panel of Fig.~\ref{fig:extinction}), suggesting the redder H$\alpha$ emitters tend to be dustier. It may be interesting to note that there is a large scatter around the $A_{\rm H\alpha}$--$M_{\star}$ relation, and that there exist some exceptionally dusty galaxies with $A_{\rm H\alpha}$$\gsim$3~mag, which cannot be identified with the simple $M_{\star}$-dependent correction\footnote{We note that there are two H$\alpha$ emitters showing $A_{\rm H\alpha}$$<0$ (i.e.\ SFR$_{\rm H\alpha}$/SFR$_{\rm UV}$$<1$). One of these sources (ID\#930) is consistent with $A_{\rm H\alpha}$$=0$ within its error bar (probably a dust-free galaxy), whilst the other source (ID\#13) shows an unrealistic negative $A_{\rm H\alpha}$ value. This is because the $B$-band flux of ID\#13 is contaminated by its close companion galaxy (due to the poorer PSF size of our $B$-band data). Although we do not remove this source from our analyses, it does not affect our conclusion at all.}. Also, there may be a weak correlation between $A_{\rm H\alpha}$ and environment for the same H$\alpha$ emitters sample (see Fig.~\ref{fig:extinction} right). Although the statistical significance is low (only $\sim$1$\sigma$ level) and there is a large uncertainty in deriving $A_{\rm H\alpha}$ of individual galaxies, this result may suggest that SF galaxies in high-density environments tend to be more highly obscured by dust (by $\sim$0.5~mag level) than field counterparts. This is qualitatively consistent with our recent study (\citealt{koy13b}), where we showed a higher dust extinction of cluster galaxies with a 24$\mu$m stacking analysis for $z=0.4$ galaxies. 

Finally, we show in Fig.~\ref{fig:SFR_density} the SFRs and specific SFRs (SSFR$=$SFR/$M_{\star}$) of all H$\alpha$ emitters as a function of local galaxy density ($\log\Sigma_{\rm 5th}$ measured with all member galaxies). The left- and right-hand panels show the results based on the two independent dust extinction correction described above. We do not find any significant SFR--density or SSFR--density correlation in any of the four panels in Fig.~\ref{fig:SFR_density}, suggesting that SF activity of SF galaxies does not significantly change with environment. By computing the Spearman's rank correlation coefficients with a sample size of $N_{\rm HAE}=43$, we cannot reject a null hypothesis that there is no significant relationship between the two variables in any of these four panels. We caution that our results are based on the purely H$\alpha$-selected galaxies (i.e. for SF galaxy population). The clear excess of non-H$\alpha$-emitting red-sequence galaxies in the very high-density region (as shown in Figs.~\ref{fig:map}--\ref{fig:colmag}) implies an anti-correlation between (S)SFR and environment, although we cannot measure SFRs of galaxies without H$\alpha$ emission.

\subsection{Star formation main sequence}

The important goal of this study is to test the environmental dependence of the SF main sequence at $z\sim 1.5$ using the H$\alpha$-selected galaxies in the newly discovered cluster field. In Fig.~\ref{fig:MS}, we show the SFR--$M_{\star}$ diagrams for H$\alpha$ emitters in the 4C\,65.22 field. The left- and right-hand panels show the result from different dust extinction correction: derived from $M_{\star}$ (left panel) and from SFR$_{\rm H\alpha}$/SFR$_{\rm UV}$ ratio (right panel). In this plot, the H$\alpha$ emitters in high-density environment ($\log\Sigma_{\rm 5th}>1.4$; top $\sim$1/3) are shown with the red circles, while the other H$\alpha$ emitters are shown with the blue circles. Regardless of the dust extinction correction procedure, we find that the location of the SF main sequence is independent of environment at $z\sim 1.5$. 

In the bottom panels of Fig.~\ref{fig:MS}, we also show the distribution of SFR offsets from the main sequence ($\Delta$MS$_{\rm offset}$) for galaxies in high-density (red filled histogram) and low-density environment (blue hatched histogram). Again, we do not see any strong difference between the two environment samples: the KS test suggests that it is unlikely that the two distributions are drawn from the same parent population in either case of the dust extinction procedure, further supporting our conclusion that the location of the SF main sequence does not significantly change with environment. One potentially interesting finding from Fig.~\ref{fig:MS} is that the H$\alpha$ emitters in the higher-density region (red symbols) tend to populate the higher envelope of the SF main sequence in the massive end (with $\log(M_{\star}/M_{\odot})>10.5$). We attempt to quantify this trend by calculating the fraction of galaxies with $\log(\Delta$MS$_{\rm offset})>0.3$ (i.e. galaxies with boosted activity). The fraction turns out to be 56$\pm$31\% and 33$\pm$19\% for high- and low-density sample, respectively. The error bars are clearly large due to the limited sample size, preventing us from drawing any firm conclusion. However, at least to say, those massive galaxies with boosted activity in high-$z$ cluster environments should be an interesting population under the influence of environmental effects.    

\begin{figure}
 \begin{center}
 \vspace{0.2cm}
 \rotatebox{270}{\includegraphics[height=8.4cm]{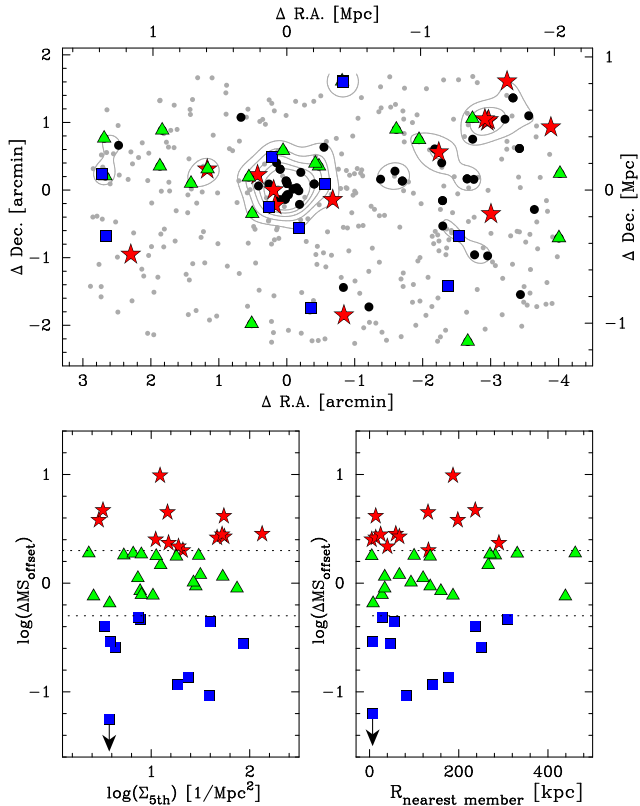}}
 \end{center}
 \vspace{-0.2cm}
\caption{ (Top): The same plot as Fig.~4, but the H$\alpha$ emitters are divided into three categories based on the offset from the main sequence ($\Delta$MS). The red stars, green triangles, and blue squares show H$\alpha$ emitters with $\log$($\Delta$MS)$>$0.3, with $-$0.3$<\log$($\Delta$MS)$<$0.3, and with $\log$($\Delta$MS)$<-$0.3, respectively. The black circles show the other member galaxies without H$\alpha$ emission. (Bottom): The $\log$($\Delta$MS) values of H$\alpha$ emitters as a function of $\log\Sigma_{\rm 5th}$ (left panel) and the distance to the nearest neighbor ($R_{\rm nearest}$; right panel). Note that the $\Delta$MS values shown in this plot are calculated with the dust extinction correction derived from SFR$_{\rm H\alpha}$/SFR$_{\rm UV}$, and the $\Sigma_{\rm 5th}$ and $R_{\rm nearest}$ are computed with all member galaxies (including non-H$\alpha$-emitters).\\
\label{fig:map_MS}}
\end{figure}

Finally, we investigate the origin of the scatter of the $\Delta$MS$_{\rm offset}$ value within our observed field. In the top panel of Fig.~\ref{fig:map_MS}, we plot the spatial distribution of the H$\alpha$ emitters by dividing the sample into three categories based on the $\Delta$MS$_{\rm offset}$ value: ``bursty'' galaxies ($\log\Delta$MS$_{\rm offset}>0.3$), ``normal'' galaxies ($-0.3<\log\Delta$MS$_{\rm offset}<0.3$), and ``semi-passive'' galaxies ($\log\Delta$MS$_{\rm offset}<0.3$). The sample is of course small, but it may be interesting to note that the north-west group located $\sim$1 Mpc away from the main body of the cluster tend to have a large number of ``bursty'' galaxies, implying that star-burst activities are most frequently triggered in poor group environments. In contrast, we can see a mix of bursty/normal/semi-passive population in the periphery of the cluster core. The local density ($\Sigma_{\rm 5th}$) of galaxies in the cluster periphery is similar to those in the north-west group, making it hard to identify the site of starburst activity with the local density approach alone (see bottom-left panel of Fig.~\ref{fig:map_MS}). 

To further investigate the environmental origin of the scatter of the SF main sequence, we show in the bottom-right panel of Fig.~\ref{fig:map_MS} the $\Delta$MS values as a function of the distance to the nearest neighboring member galaxy ($R_{\rm nearest}$). In this plot, we find a tentative hint that galaxies having a close companion (with $\lsim$200~kpc) tend to show a larger scatter around the main sequence, whilst isolated galaxies tend to be located on the main sequence. The sample is small, but this plot may suggest a potential link between the galaxy--galaxy encounter and the deviation around the SF main sequence---i.e.\ a galaxy tends to show a boosted/truncated SF activity once they become a satellite of another galaxy (c.f.\ \citealt{par09}; \citealt{hwa11}). However, because of the mixture of bursty/normal/passive population, we cannot detect an environmental variation in the location (or zero point) of the SF main sequence.

\section{DISCUSSION}

\subsection{Galaxy population in $z \gsim 1.5$ clusters}

It has been a long observational challenge to find a galaxy cluster at $z\gsim 1.5$ until recently. However, an increasing number of studies have identified clusters at such frontier redshifts with various techniques (e.g. \citealt{kur09}; \citealt{san11}; \citealt{nas11}; \citealt{sta12}; \citealt{muz13}), and the nature of such high-$z$ clusters and their member galaxies are now being investigated. Some studies point out that SF activity in high-$z$ cluster cores are comparable to (or even higher than) general field galaxies (e.g.\ \citealt{hay10}; \citealt{hil10}; \citealt{tra10}; \citealt{tad12}); hence galaxy clusters at these redshifts are interpreted as being in the ``transitional phase'' or final quenching period (\citealt{hay11}; \citealt{bro13}; \citealt{sma14}; \citealt{alb14}). 

Interestingly, our newly discovered cluster reported in this paper resembles more like $z\lsim 1$ clusters, characterized by a strong concentration of passive red galaxies in the cluster center and star-forming population surrounding the core (e.g.\ \citealt{koy10}; 2011). Our finding of the suppressed SF activity in the cluster core is also similar to the situation in the well-studied rich cluster XMMU\,J2235.3$-$2557 at $z=1.39$ (\citealt{lid08}, \citealt{bau11a}, \citealt{gru12}; \citealt{san13}). Recently, \cite{tan13} studied an X-ray detected $z=1.6$ group in Chandra Deep Field South. Based on the SED and morphological analyses using the CANDELS data, these authors suggest that most galaxies in this $z\sim 1.6$ group are quiescent early-type systems (with a few of them specroscopically confirmed). The size of the spatial extension of quiescent galaxies in the system reported by \cite{tan13} is also qualitatively consistent with that of our newly discovered system at $z=1.5$ ($\sim$200--300~kpc scale). 

In our current work, by utilizing the narrow-band H$\alpha$ imaging technique, we find a clear ``deficit'' of blue galaxies in the central cluster region ($r_c$$<$250~kpc). We do find a few H$\alpha$ emitters within the $r_c=$250~kpc circle, but those H$\alpha$ emitting galaxies in the cluster core show redder colors than typical blue H$\alpha$ emitters seen in general field environment (\S~4.2). We anticipate that the clear lack of low-mass blue galaxies in the cluster central region is due to the rapid SF quenching and rapid mass growth triggered by environmental effects. If a low-mass blue galaxy enters a rich cluster environment, its SF activity will soon be quenched. This quenching mechanism should accompany an intense short-lived starburst or merging event, because the red-sequence galaxies dominating the cluster core tend to be much more massive than typical blue SF galaxies in the surrounding field.

\subsection{Color--density, SFR--density relation at $z>1$}

The color--density relation has been investigated out to $z\sim 1$ or even beyond (e.g.\ \citealt{kod01}; \citealt{cuc06}; \citealt{coo10}). Our data suggest that the color--density relation is already in place around the radio galaxy at $z\sim 1.5$. However, the correlation becomes much weaker once we remove non-H$\alpha$-emitting galaxies, suggesting that the color--density relation is a product of passive red-sequence galaxies dominating the highest-density cluster core. A similar result was drawn from morphological analysis of cluster/field galaxies at $z\sim 1.6$ by \cite{bas13}. These authors showed that the morphologies of SF galaxies in cluster and field environments at $z\sim 1.6$ show no significant difference. We note, however, that our data covers only a limited FoV (2.0$\times$3.6 Mpc$^2$), and it is possible that we do not sample galaxies in very low-density environments. We also note that the definition of the H$\alpha$ emitters in this study is EW$_{\rm rest}$(H$\alpha$$+$[N{\sc ii}])$\gsim$20\AA, which might be too large if we consider local SF galaxies (e.g.\ \citealt{ken83}). At $z\sim 1.5$, however, it is expected that the EW$_{\rm rest}$(H$\alpha$$+$[N{\sc ii}])$\gsim$20~\AA\ criterion allows us to select a fairly complete sample of normal main-sequence galaxies. Therefore we do not expect that the EW cut does not make any strong impacts on our results, but we need to keep in mind that our survey can potentially miss low-EW (but still star-forming) galaxies. This is an inevitable problem associated with any NB-based study, and we need future deep imaging/spectroscopic observations to construct a more complete sample of galaxies down to such less active population.

It is worth mentioning that recent studies on $z>2$ proto-clusters suggest that SF galaxies in higher-density environments tend to have redder colors, implying a color--density correlation {\it amongst star-forming galaxies} (see \citealt{koy13a}; \citealt{hay12}). \cite{koy13a} showed that most of H$\alpha$ emitters with red $J-K$ colors in the core of the PKS1138--262 proto-cluster at $z=2.2$ are very massive system (with $\log M_{\star} \gsim 10^{11}M_{\odot}$). Such an accelerated galaxy stellar mass growth is also suggested by some recent studies on $z\gsim 2$ proto-clusters (\citealt{ste05}; \citealt{hat11}; \citealt{mat11}; \citealt{coo14}). In our current analysis, we find a number of red massive galaxies {\it without star formation} in the core of the $z\sim 1.5$ cluster. The stellar masses of these passive population are comparable to those of massive SF galaxies in the $z>2$ proto-cluster enviromnents. If we assume that those massive SF galaxies reported in $z>2$ proto-cluster environments are the direct progenitors of passive red galaxies seen in the $z\sim 1.5$ cluster core, they are expected to have stopped their SF activity within a relatively short time interval between $z\sim 2$ and $z\sim 1.5$. 

There have been a lot of debates on the ``reversal'' of the SFR--density relation in the distant universe. \cite{elb07} first noted this possibility with the multi-wavelength dataset in the GOODS fields (see also \citealt{coo08}; \citealt{ide09}). This reversal of the SF--density relation is still under debate, and it is likely that the results could be uncertain depending on the sample definitions and/or the definitions of environment (\citealt{sob11}; \citealt{pat11}; \citealt{sco13}; \citealt{zip13}). At least for star-forming galaxies, recent studies have obtained a similar conclusion that there is no significant SFR--density or SSFR--density correlation amongst SF galaxies (\citealt{mcg11}; \citealt{pen10}; 2012; \citealt{tad12}; \citealt{muz12}). The results drawn from our current work is consistent with those recent studies. We caution that the ``flat'' SFR--density relation may not hold at $z\gsim 2$. As we demonstrated in \cite{koy13a}, SF galaxies in proto-cluster environments tend to be more massive than those in underdense regions. Although most of the H$\alpha$ emitters in proto-cluster environments tend to be located on the SF main sequence, they are skewed to the massive end of the main sequence; hence their SFRs tend to be higher (on average) than those in general field environments.

\subsection{Environmental impacts on the SF main sequence}

A vital approach to understanding the environmental effects on SF galaxies is to examine the environmental impacts on the SF main sequence. Recent studies have investigated the SF main sequence mainly for distant {\it field} galaxies. Those studies have confirmed the existence of the SF main sequence out to $z\sim 2.5$ or above, but the environmental dependence of SF main sequence is much poorly understood due to the lack of well-defined SF galaxy samples in clusters/groups in the distant unverse. 

In \cite{koy13b}, we studied the environmental dependence of the SF main sequence using our purely H$\alpha$-selected galaxy samples at $z=0.4, 0.8, 2.2$ in distant cluster environments and general field environments. We found that there is no significant environmental variation in the SF main sequence out to $z\sim 2$ (with a difference of $\sim$0.2~dex at maximum). In that work, we used $M_{\star}$-dependent extinction correction for all H$\alpha$ emitters, which is reported to be unchanged (at least on average) throughout the cosmic time (\citealt{sob12}). However, we need to test the validity of this procedure by comparing the results derived with different methods of dust extinction correction. In our current analysis, we test this concern by estimating the H$\alpha$ dust extinction in a more realistic way using the SFR$_{\rm H\alpha}$/SFR$_{\rm UV}$ ratio. However, we again find that the location of the SF main sequence does not change with environment (with a difference of $\lsim$0.1~dex level), confirming the conclusion drawn by \cite{koy13b}. 

We caution, however, that the dust extinction issue is still an important key question to fully assess the real enviromental impacts on the SF main sequence. Indeed, we do find a marginal trend that the dust extinction effect ($A_{\rm H\alpha}$) is weakly correlated with local density in the sense that SF galaxies in higher-density environment tend to be more highly obscured by dust. A similar implication was obtained for $z=0.4$ galaxies by \cite{koy13b}. In that work, we used SFR$_{\rm IR}$/SFR$_{\rm H\alpha}$ ratio (with 24$\mu$m stacking analysis) to show that SF galaxies in high-density environments tend to be dustier (by $\sim$0.5~mag level) than those in underdense environments, in qualitative agreement with the current work. 

Our studies thus suggest a possibility that SF galaxies surviving in high-$z$ cluster environments tend to be dustier, and perhaps the nature/mode of SF activity in cluster environments may be different from those in general field environments. We recall that the environmental impacts on the dust properties of galaxies are still very pooly understood at this moment, and there are some contradicting results on this issue (\citealt{gar10}; \citealt{pat11}). Therefore, a special care must be taken on the dust extinction correction when we interprete the environmental dependence of the SF main sequence, and this key issue needs to be studied more in detail with larger galaxy samples in the future. 

Finally, we comment on the possibility of the environmental impacts on the ``scatter'' of the SF main sequence. As we showed in Fig.~\ref{fig:map_MS}, we do not find a correlation between $\Delta$MS and $\Sigma_{\rm 5th}$, consistent with the environmental independence of the location of the main sequence. On the other hand, we find a tentative hint that the $\Delta$MS value more strongly deviates for those having a close companion within $\lsim$200~kpc, while more isolated galaxies tend to be more steadily located on the SF main sequence. The sample is clearly too small to draw any firm conclusion, but this result may provide us with a new insight on the origin of the scatter of the SF main sequence. If a galaxy approaches another galaxy, their SF activity would be disturbed (some are boosted and others are truncated) due to the galaxy-galaxy interaction or halo gas stripping (satellite quenching). This would explain the larger scatter of the $\Delta$MS for those with small $R_{\rm nearest}$, and the apparent environmental independence of the location of the SF main sequence.

\section{Summary and Conclusions}

We performed a broad-band and narrow-band (H$\alpha$) imaging survey of a radio galaxy field (4C\,65.22) at $z=1.52$ over the 7$'\times$4$'$ FoV (corresponding to 3.6$\times$2.0 Mpc$^2$) with MOIRCS/Suprime-Cam on the Subaru Telescope. Based on this new dataset, we find a rich cluster candidate around the radio galaxy. We also studied an environmental dependence of galaxy properties around this newly discovered structure at $z\sim 1.5$. Our findings are summarized as the following.

(1) With the photo-$z$ and H$\alpha$ emitter selections, we discovered a strong over-density of galaxies around 4C\,65.22. We find that the cluster central region ($r$$\lsim$250~kpc from the density peak) is clearly dominated by passive red-sequence galaxies without H$\alpha$ emission, while the H$\alpha$-emitting galaxies are preferentially located in the cluster outskirts, showing a sharp decline in the H$\alpha$ emitter fraction toward the cluster center. This spatial segregation is similar to that seen in lower redshift clusters, suggesting that the newly discovered structure is a well-matured system. 

(2) The color--density and $M_{\star}$--density relations are already in place at $z\sim 1.5$. These environmental trends are mostly driven by passive red/massive galaxies residing in the cluster central region, whilst such environmental trends become much weaker when we consider only H$\alpha$ emitters. This is also the case for SFR--density and SSFR--density relation. An excess of non-H$\alpha$-emitting red-sequence galaxies in the cluster core strongly suggests that the SF activity is suppressed in the very rich environment at $z=1.5$, but the SF activity amongst SF galaxies are almost independent of environment. 

(3) There may exist a weak correlation between dust attenuation (derived with SFR$_{\rm H\alpha}$/SFR$_{\rm UV}$ ratio) and local environment for SF galaxies at $z=1.5$: galaxies in high-density environments tend to be dustier by $\sim$0.5--1.0 mag level. However, even if we take this point into account, we cannot see a detectable environmental variation in the location of the SF main sequence, consistent with recent studies. We do not find any correlation between the SFR offset from the SF main sequence ($\Delta$MS) and $\Sigma_{\rm 5th}$, but we find a tentative hint that galaxies having a close companion (at a moderate distance of $\lsim$200~kpc) tend to be more largely scattered from the SF main sequence than more isolated galaxies.

\acknowledgments

The broad-band and narrow-band imaging data used in this paper are 
collected at the Subaru Telescope, which is operated by the National 
Astronomical Observatory of Japan (NAOJ). We thank 
the referee for reviewing our paper and providing us with useful comments
which improved the paper. Y.K., K.T., and M.H. acknowledge support from 
the Japan Society for the Promotion of Science (JSPS) through 
JSPS research fellowships for Young Scientists. 
This work was financially supported in part by a Grant-in-Aid for the
Scientific Research (Nos.\,21340045; 24244015) by the Japanese 
Ministry of Education, Culture, Sports and Science. \\

{\it Facilities:} \facility{Subaru} .

\end{document}